\documentclass[a4paper,10pt,authoryear,preprint,review,square,comma]{submission}

\allowdisplaybreaks[4]

\begin{document}
\setcounter{page}{1}

\ensubject{subject}

\ArticleType{Article}
\Year{2017}
\Vol{1}
\No{1}

\title{
Relativistic Difference of LIGO Signal}{Relativistic Difference of LIGO Signal}
\author[]{Rui Chen}{}

\AuthorMark{Rui Chen}

\address[]{Shenzhen Foreign Languages School, Shenzhen, China}


\abstract{Signal waves of the monotonously increasing frequency detected by LIGO are generally considered to be gravitational waves of spiral binary stars, thus confirming the general theory of relativity. Here we present a universal method for signal wave spectrum analysis, introducing the true conclusions of numerical calculation and image analysis of GW150914 signal wave. Firstly, numerical calculation results of GW150914 signal wave frequency change rate obey the com quantization law which needs to be accurately described by integers, and there is an irreconcilable difference between the results and the generalized relativistic frequency equation of the gravitational wave. Secondly, the assignment of the frequency and frequency change rate of GW10914 signal wave to the generalized relativistic frequency equation of gravitational wave constructs a non-linear equation group about the mass of wave source, and the computer image solution shows that the equation group has no GW10914 signal wave solution. Thirdly, it is not unique to calculate the chirp mass of the wave source from the different frequencies and change rates of the numerical relativistic waveform of the GW150914 signal wave, and the numerical relativistic waveform of the GW150914 signal wave deviates too far from the original waveform actually. Other LIGO signal waveforms do not have obvious characteristics of gravitational frequency variation of spiral binary stars and lack precise data, so they cannot be used for numerical analysis and image solution. Therefore, the generalized relativistic frequency equation of gravitational wave has not been verified by LIGO signal wave up to now. Does this mean that the gravitational wave signals from some spiral binaries that may be detected in future obey the same com quantization law? The answer remains unclear.}


\maketitle

\renewcommand{\thefootnote}{\fnsymbol{footnote}}

\normalsize
\begin{multicols}{2}
\section{Introduction}
In 1915, Einstein establishes the equation of gravitational field\cite{Einstein:1915} and
founded general relativity\cite{Einstein:1916}. In 1916, Einstein predicted
the existence of gravitational waves based on general theory of relativity,
and published the first paper on gravitational waves\cite{Einstein:1917}.
According to the general theory of relativity, when an object accelerates,
it will generate gravitational radiation and escape from the gravitational
field source and propagate in the vacuum to form gravitational waves, which
is considered that accelerated masses stimulate fluctuations in space-time.
In fact, gravitational waves are fluctuating gravitational fields. In 1916,
Schwarzschild calculated the static spherical symmetric
solution\cite{Schwarzschild:1916,Schwarzschild:1917} of the
Einstein field equation, and the singularity in this particular solution was
interpreted as the radius of the black hole's horizon. In February 1918,
Einstein published the second paper on gravitational waves and gave formulas
for the calculation of gravitational radiant energy\cite{Einstein:1918}. In
1963, Kerr found the solution of the rotating black hole\cite{Kerr:1963}
for the field equation.

Since Einstein proposed the concept of gravitational waves, the theoretical
and experimental detection principles of gravitational
waves\cite{Weber:1961,Abramovici:1992,Harry:2010}
have been continuously developed within the framework of general relativity.
Although gravitational waves exist widely, they are generally difficult to
detect due to their very weak energy. According to theoretical predictions,
massive binary black holes or binary compact stars can generate powerful
energy gravitational waves when merger, so compact binary stars like binary
black holes become an important model for the study of gravitational wave
theory and experimental detection principles. The main vibration frequency
and energy of the gravitational waves generated by the binary star during
the process of inspiral, merger and ringdown first increase and then
decrease with time\cite{Baker:2001,Damour:2008}. In 1974,
Hulse and Taylor used a radio telescope to find a pulsed binary neutron star
that rotates one revolution every 8 hours\cite{Hulse:1974}. The radiation
of gravitational waves in the process of binary satellite spiralling reduces
the system energy and decreases the revolution period. Precise measurement
results show that the period of revolution of the twin neutron star is
reduced by $10^{-4}$ second per year, which is in accordance with the
theoretical value. The discovery of the pulsed binary neutron star is
considered to prove indirectly the existence of the gravitational wave. In
1995, Blanchet et al. derived the Blanchet frequency
equation\cite{Blanchet:1995} of gravitational wave from the binary star
wave source, which is one of the important corollaries of the general
relativistic gravitational wave theory.

In February 2016, LIGO's two detectors at Livingston and
Hanford received the signal wave named GW150914\cite{Abbott:2016} at 6.9 ms
intervals on September 14, 2015. In the signal waveforms, the vibration curve with the
strain not exceeding about $1.2\times 10^{-21}$ for a period of time from
0.25 s to 0.45 s is considered as the gravitational wave of the spiral
binary black holes, which was generated by the merger of two black holes
whose mass was $29_{-4}^{+4} M_\odot $ and $36_{-4}^{+5} M_\odot $
respectively from the solar system 1.5 billion light years away, 1.3 billion
years ago, and the mass of the merged black hole is $62_{-4}^{+4} M_\odot
$\cite{Abbott:2017}. In less than two years after that, LIGO successively
announced the detection of four gravitational wave signals
GW151226\cite{Abbott:2018}, GW170104\cite{Scientific:2017},
GW170814\cite{Abbott:2019}, and GW170817\cite{Abbott:2020} from the spiral
binary holes or spiral binary neutron stars. The widely accepted conclusion is that these signals are gravitational waves generated by the merging of spiral binary black holes or spiral binary neutron stars. They constitute the last piece of Einstein's general relativity. Not only does the
black hole predicted by general relativity exist, but the binary black holes
also merge frequently, even though space- time is ancient-far. Why do all
the gravitational wave signals detect by LIGO come from the merger of
ancient-far binary black holes or binary neutron stars? How is the detailed
frequency law of the gravitational wave from binary stars? Scientific
assertions require scientific argumentation. The vibration curve of signal
wave of GW150914 is clear, and the part where the frequency changes
monotonically accords with the characteristics of spiral binaries. However,
the frequency distribution of signals such as GW151226, GW170104, GW170814,
GW170817 does not have precise rules, and quantitative calculations cannot
be made to arrive at reliable conclusions, and there is actually a lot of
uncertainty. Therefore, at present, the GW150914 signal wave is still the only experimental basis for identifying gravitational wave and testing gravitational wave theory.

Whether a signal wave is a gravitational wave can be judged by comparing the results of numerical calculation and image method with the inferences of the theoretical model. Here, data of the GW150914 signal wave is analyzed in detail. Firstly, time of main strain peak is extracted from GW150914 signal wave database, and the time of wide and uncertain peak is corrected within the error range to determine the frequency of the main strain peak. Then, the Lagrange frequency change rate, which represents the average frequency change rate, is calculated. It is proved that the frequency change of GW1500914 signal wave presents a generalized quantization law called com quantum law, which needs to be described by integers, and there is a great difference between frequency distribution of the GW150914 signal wave and the general relativistic Blanchet frequency equation of gravitational wave. Thirdly, the image method is used to compare the average frequency change rate curve of the GW150914 signal wave with Blanchet frequency equation function curve, which further proves that there is a great difference between the frequency change law of the GW150914 signal wave and Blanchet frequency equation. Finally, the frequency change rate of LIGO's numerical relativistic gravitational waveform is calculated, and the result also does not conform to the general relativistic Blanchet frequency equation.

\section{Frequency distribution and change laws of GW150914 signal wave}

The frequency of the main vibration part of the GW150914 signal wave
increases monotonically, and the strain of the main vibration part shows a
tendency to change synchronously with the frequency, but does not increase
strictly monotonously, which is mainly caused by noise. The detection of
gravitational waves usually uses filtering techniques to shield the noise.
It should be pointed out that the filtering technology can only shield the
noise of the expected frequency distribution, and the isolated wave or the
noise of unintended frequency distribution can still reach the detector. The
noise with similar energy mixed in the gravitational wave strengthens or
weakens the strain at a certain moment, so that the gravitational waveform
is distorted to varying degrees. The gravitational wave waveform is also
distorted by the forced vibration of the detection instrument due to the
tremor of the crust without a fixed frequency. Therefore, the strain
corresponding to the gravitational wave signal may be shifted or masked.
Strain distribution of the GW150914 signal wave is more complex than its
frequency distribution. Here we mainly analyze its frequency distribution
and its variation laws.

\begin{figure}[H]
\centerline{\includegraphics[width=\linewidth]{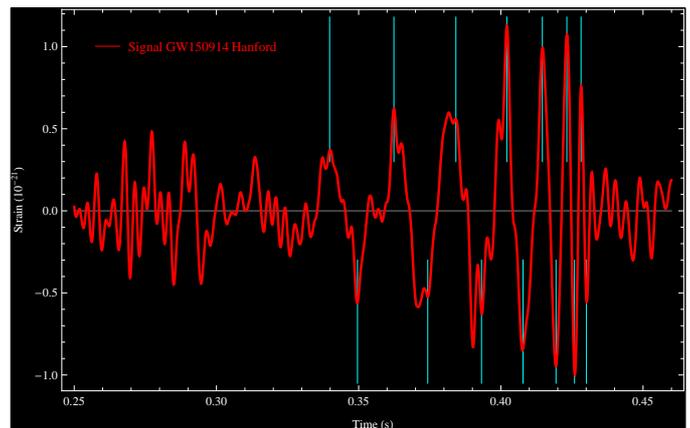}}
\caption{The positive and negative strain time of the GW150914 signal wave
\cite{https:1}.}
\label{fig1}
\end{figure}

As shown in Figure 1, the main positive and negative strain peak time are
extracted from the Hanford database\cite{https:1} of the GW150914 signal wave
and marked by vertical lines in reverse time sequence. The rightmost
vertical line corresponds to the sequence number 1. Correct the time of wide
or uncertain peak within the allowable range of error. The accuracy of recording
time in LIGO database reaches $10^{-9}\mbox{s}$. The correction time of wide
or uncertain peaks are obtained by using the characteristic equation
correction method and the accuracy of $10^{-9}\mbox{s}$ is naturally retained.
The formulas for the period and frequency are $T_i =t_i -t_{i+1} $ and $f_i =1 \mathord{\left/
{\vphantom {1 {T_i }}} \right. \kern-\nulldelimiterspace} {T_i }$,
respectively. The time of the positive and negative strains, the
corresponding period, and the frequency calculation result are listed in
Table 1 in reversed time order. The positive strain peaks have 6 main
frequencies from 36.55320819Hz to 197.3975904Hz; the negative strain peaks
have a total of seven main frequencies from 35.37953557Hz to 230.7600891Hz.
The period and frequency of the strain peaks characterize the period and
frequency of the signal wave.

As shown in Figure 2, according to the positive and negative strain time
$t_i $ and frequency $f_i $ listed in Table 1, the frequency and time curves
of the positive and negative strains of the GW150914 signal wave are plotted
respectively. The two polylines have the same monotonous change trend, and
they are very close to each other. A single polyline that plots the
frequency and time of the positive and negative strain mixtures reflects the
fluctuation characteristics of the frequency distribution.
\begin{figure}[H]
\centerline{\includegraphics[width=\linewidth]{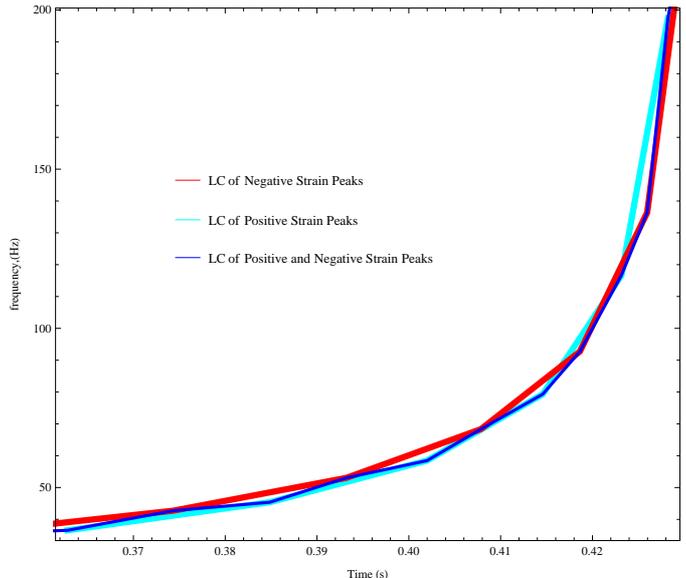}}
\caption{Frequency-time polylines of strain peaks of the GW150914 signal wave.}
\label{fig2}
\end{figure}

\end{multicols}
\begin{center}
\renewcommand\arraystretch{0.9}
\begin{table}[htbp]\footnotesize \scriptsize \tiny
\begin{center}
\caption{Positive and negative strain peak times of the GW150914 signal wave and its frequency distribution}
\resizebox{\textwidth}{!}{
\begin{tabular}{c|ccc|ccc}
\toprule
\raisebox{-1.50ex}[0cm][0cm]{$i$ }&
\multicolumn{3}{c|}{Positive strain observation values} &
\multicolumn{3}{c}{Negative strain observation values}  \\
\cline{2-7}
&
$t_i \mbox{ /s}$&
$T_i \;\mbox{/s}$&
$f_i \;\mbox{/Hz}$&
$t_i \mbox{ /s}$&
$T_i \;\mbox{/s}$&
$f_i \;\mbox{/Hz}$ \\
\hline
1&
0.428222656&
0.005065918&
197.3975904&
0.430297903&
0.004333505&
230.7600891 \\
2&
0.423156738&
0.008573092&
116.6440307&
0.425964398&
0.007333624&
136.3582345 \\
3&
0.414583646&
0.012607488&
79.31794085&
0.418630774&
0.010784741&
92.72359945 \\
4&
0.401976158&
0.017110162&
58.44479852&
0.407846033&
0.014636434&
68.32265222 \\
5&
0.384865996&
0.022037889&
45.37639637&
0.393209599&
0.018851727&
53.04553744 \\
6&
0.362828106&
0.027357380&
36.55320819&
0.374357872&
0.023402144&
42.73112738 \\
7&
0.335470727&
&
&
0.350955728&
0.028264927&
35.37953557 \\
8&
&
&
&
0.322690801&
&
 \\
\hline
\end{tabular}
}
\label{tab1}
\end{center}
\end{table}
\end{center}
\begin{multicols}{2}
\normalsize

It is assumed that the signal wave of GW150914 is the gravitational wave
of a helical double black hole. In order to determine the chirp mass of
GW150914 wave source, LIGO uses the low frequency approximation of the highly
nonlinear Blanchet frequency equation of general relativity.
In fact, the high and low
frequencies are relative and there is no clear demarcation. To judge whether
the frequency distribution and variation laws of a signal wave accord with
the Blanchet frequency equation, it is necessary to calculate the time
derivative of the frequency. The frequencies of the GW150914 signal wave and
their change rates are all discrete. The classical method of calculation is
to first modify the graph of frequency change over time into a smooth curve,
draw the tangent of the position of each frequency, and then measure the
slope of these tangents. Therefore, the derivative of the frequency versus
time $\dot {f}_i $ at each frequency is obtained. However, the correction of
a polyline to a smooth curve has great uncertainty, and the tangent line
also has uncertainty. The calculation of the rate of change is actually
uncertain. As we all know, the mathematical significance of the average rate
of change of the discrete variation is consistent with the Lagrange mean
value theorem\cite{Sahoo:1998}, so the Lagrange mean value theorem can be
used to calculate the time derivative of the frequency of the gravitational
wave, that is, the Lagrange frequency change rate. The integer $i$ in a
table is an inverse time sequence, and the variation rate of discrete
frequency conforming to the meaning of Lagrange mean value theorem is called
Lagrange frequency change rate. It is defined as,

\begin{equation}
\label{eq1}
{\dot f_i} = \frac{{{f_{i - 1}} - {f_{i + 1}}}}{{{T_i} + {T_{i - 1}}}}
\end{equation}

The Lagrange change rate has a definite value for calculating the average change rate of the discrete frequency, which is better than the tangent slope measured after the frequency time polyline is corrected to a smooth curve, because the correction of the curve and the measurement results of the slope are uncertain.

The column in Table 2 that $\dot {f}_i $ is located
lists the Lagrange frequency change rates of the main positive and negative
strain peaks of the GW150914 signal wave. According to dimension analysis,
the time derivative of the frequency should be represented by the square of
the frequency. The column in Table 2 where $\dot {f}_i \dot {f}_i^{-2} $ is
located also lists the ratio of the frequency change rate of the positive
and negative strains of the GW150914 signal wave to their square of the
frequency. Numerical results show that the frequency distribution and
changes in the positive and negative strain peaks are different, but the
ratio $\dot {f}_i \dot {f}_i^{-2} $ has the same distribution with high
accuracy, which is expressed as the equation of following form,
\begin{equation}
\label{eq2}
\left\{ {\begin{gathered}
 \dot {f}_2^+ =0.636307692f_2^2 \\
 \dot {f}_3^+ =0.436753649f_3^2 \\
 \dot {f}_4^+ =0.334368530f_4^2 \\
 \dot {f}_5^+ =0.271585859f_5^2 \\
 \vdots \\
 \end{gathered}} \right\}\Leftrightarrow \left\{ {\begin{gathered}
 \dot {f}_2^- =0.636307692f_2^2 \\
 \dot {f}_3^- =0.436753649f_3^2 \\
 \dot {f}_4^- =0.334368530f_4^2 \\
 \dot {f}_5^- =0.271585859f_5^2 \\
 \dot {f}_6^- =0.228972362f_6^2 \\
 \end{gathered}} \right\}
\end{equation}
The positive and negative superscripts represent the positive and negative
strains, respectively, respectively. Discrete laws of physical quantities need to be described by integers. Discrete laws are essentially generalized quantization laws including quantization laws, which are called com quantum laws. Equation (\ref{eq2}) shows that the discrete frequencies of the GW150914 signal wave imply a generalized quantization law closely related to quantum numbers in accordance with the dimensional law. This opens the prelude to the gravitational com quantum theory, which systematically describes the quantization law of gravitational systems.  It also means that the GW150914 gravitational wave does not correspond to the Blanchet frequency equation of general relativity.

\end{multicols}
\begin{center}
\renewcommand\arraystretch{0.9}
\begin{table}[htbp]
\begin{center}
\caption{Relativistic differences of frequency change rule of the GW150914 signal wave}
\resizebox{0.98\textwidth}{!}{\scriptsize 
\begin{tabular}{c|cccc|cccc}
\toprule
\raisebox{-1.50ex}[0cm][0cm]{$i$}&
\multicolumn{4}{c|}{Positive strain observation values} &
\multicolumn{4}{c}{Negative strain observation values}  \\
\cline{2-9}
 &
$f_i $&
$\dot {f}_i $&
${\dot {f}_i } \mathord{\left/ {\vphantom {{\dot {f}_i } {f_i^2 }}} \right. \kern-\nulldelimiterspace} {f_i^2 }$&
${\dot {f}_i } \mathord{\left/ {\vphantom {{\dot {f}_i } {f_i^{{11} \mathord{\left/ {\vphantom {{11} 3}} \right. \kern-\nulldelimiterspace} 3} }}} \right. \kern-\nulldelimiterspace} {f_i^{{11} \mathord{\left/ {\vphantom {{11} 3}} \right. \kern-\nulldelimiterspace} 3} }$&
$f_i $&
$\dot {f}_i $&
${\dot {f}_i } \mathord{\left/ {\vphantom {{\dot {f}_i } {f_i^2 }}} \right. \kern-\nulldelimiterspace} {f_i^2 }$&
${\dot {f}_i } \mathord{\left/ {\vphantom {{\dot {f}_i } {f_i^{{11} \mathord{\left/ {\vphantom {{11} 3}} \right. \kern-\nulldelimiterspace} 3} }}} \right. \kern-\nulldelimiterspace} {f_i^{{11} \mathord{\left/ {\vphantom {{11} 3}} \right. \kern-\nulldelimiterspace} 3} }$ \\
\hline
1&
197.3975904&
&
&
&
230.7600891&
&
&
 \\
2&
116.6440307&
8657.49422&
0.636307692&
0.000228505&
136.3582345&
11831.23042&
0.636307692&
0.000176143 \\
3&
79.31794085&
2747.763841&
0.436753649&
0.000298275&
92.72359945&
3755.061951&
0.436753649&
0.000229925 \\
4&
58.44479852&
1142.134177&
0.33436853&
0.000379882&
68.32265222&
1560.827218&
0.33436853&
0.000292831 \\
5&
45.37639637&
559.1999942&
0.271585859&
0.000470461&
53.04553744&
764.1961764&
0.271585859&
0.000362654 \\
6&
36.55320819&
&
&
&
42.73112738&
418.0919129&
0.228972362&
0.000438405 \\
7&
&
&
&
&
35.37953557&
&
&
\\
\hline
\end{tabular}
}
\label{tab2}
\end{center}
\end{table}
\end{center}
\begin{multicols}{2}

\section{Numerical proof of relativistic difference of GW150914 signal wave}

The derivation of the general relativity theory that describes the laws of
frequency distribution and variation of gravitational waves is a highly
nonlinear Blanchet frequency equation\cite{Blanchet:1995}, which cannot be
accurately solved at present. Because the frequency and strain of the signal wave increase monotonously, LIGO inferred that the GW150914 signal wave is the gravitational wave from a merge of a pair of spiral binary black holes, and took the zero order approximation of the Blanchet frequency equation at low frequencie\cite{Abbott:2016},
\begin{equation}
\label{eq3}
\dot {f}=\frac{96\pi ^{8 \mathord{\left/ {\vphantom {8 3}} \right.
\kern-\nulldelimiterspace} 3}G^{5 \mathord{\left/ {\vphantom {5 3}} \right.
\kern-\nulldelimiterspace} 3}m_1 m_2 }{5c^5\left( {m_1 +m_2 } \right)^{1
\mathord{\left/ {\vphantom {1 3}} \right. \kern-\nulldelimiterspace}
3}}f^{{11} \mathord{\left/ {\vphantom {{11} 3}} \right.
\kern-\nulldelimiterspace} 3}
\end{equation}

Among them, the universal gravitational constant is $G=6.674\times
10^{-11}\mbox{m}^3kg^{-1}s^{-2}$, the speed of light in the vacuum is
$c=2.998\times 10^8\mbox{m s}^{-1}$, and $m_1 $ and $m_1 $ is the masses of
two black holes in binary black hole's gravitational wave source
respectively. However, low frequency approximation is not a scientific method.
Because low-frequency and high-frequency are only relative, there is no clear
boundary between them. Theoretically, the frequency conversion motion
has a low frequency approaching to zero. Low frequency approximation (3)
makes it easy for readers to shift their attention to the so-called chirp mass,
while ignoring the difference between the frequency distribution of the GW150914
signal wave and the Blanchet frequency equation.

Note that the approximate
theoretical value $\dot {f}\propto f^{{11} \mathord{\left/ {\vphantom {{11}
3}} \right. \kern-\nulldelimiterspace} 3}$ derived from the formula (\ref{eq3}) does
not conform to the observed value (\ref{eq2}) clearly. The low-frequency
approximation of the Blanchet frequency equation requires $\dot {f}_i \dot
{f}_i^{{-11} \mathord{\left/ {\vphantom {{-11} 3}} \right.
\kern-\nulldelimiterspace} 3} $ to be an approximate constant, but the $\dot
{f}_i \dot {f}_i^{{-11} \mathord{\left/ {\vphantom {{-11} 3}} \right.
\kern-\nulldelimiterspace} 3} $ value of the GW150914 signal wave listed in
Table 2 varies with frequency, which is a prominent contradiction,
manifesting as that the first significant digits of the $\dot {f}_i \dot
{f}_i^{{-11} \mathord{\left/ {\vphantom {{-11} 3}} \right.
\kern-\nulldelimiterspace} 3} $ values corresponding to each frequency of
positive and negative strain peaks are quite different. This difference
cannot be eliminated by a numerical method that corrects the strain peak
time or redefines the rate of change of the discrete frequency. Although
Equation (\ref{eq3}) is the zero-order approximation of the Blanchet's frequency
equation at low frequencies, the zero-order approximation embodies the main
rule of the Blanchet's frequency equation. The high-order approximation of
the Blanchet frequency equation has a small effect on the first significant
figure, and it cannot change the conclusion that the $\mathop {\dot f}\nolimits_i \mathop {\dot f}\nolimits_i^{ - {{11} \mathord{\left/ {\vphantom {{11} 3}} \right.\kern-\nulldelimiterspace} 3}} $ values of the GW150914
signal wave are not constant and it are inconsistent with the Blanchet
equation.

The amplitudes and orbital contraction rates of the general relativistic
quadrupole moments of binary stars with unequal masses are related to the
system's mass, which happens to be in the zero-order approximation (\ref{eq3}) of
the Blanchet's frequency equation,
\begin{equation}
\label{eq4}
\mathscr{M}=\frac{\left( {m_1 m_2 } \right)^{3 \mathord{\left/ {\vphantom {3 5}}
\right. \kern-\nulldelimiterspace} 5}}{\left( {m_1 +m_2 } \right)^{1
\mathord{\left/ {\vphantom {1 5}} \right. \kern-\nulldelimiterspace}
5}}=\frac{c^3}{G}\left( {\frac{5}{96}\pi ^{{-8} \mathord{\left/ {\vphantom
{{-8} 3}} \right. \kern-\nulldelimiterspace} 3}f^{{-11} \mathord{\left/
{\vphantom {{-11} 3}} \right. \kern-\nulldelimiterspace} 3}\dot {f}}
\right)^{3 \mathord{\left/ {\vphantom {3 5}} \right.
\kern-\nulldelimiterspace} 5}
\end{equation}

According to the report of LIGO, the masses of the two black holes of the
GW150914 signal source are $29_{-4}^{+4} M_\odot $ and $36_{-4}^{+5} M_\odot
$, respectively, where $M_\odot $ represents the mass of the sun. From this,
the estimated value of the chirp mass of the wave is
\[
\mathscr{M}_{\mbox{LIGO}} =28.10_{-3.51}^{+3.89} M_\odot
\]

If the GW150914 signal wave is the gravitational wave radiated by a pair of spiral binary stars, the results of chirp mass calculated by different frequencies of the gravitational waves should be approximately equal, because the chirp mass of a spiral binary star is unique.

However, according to the lower frequency $f_5 =45.37639637\mbox{Hz}$ of the
positive strain of the GW150914 signal wave and its Lagrange change rate
$\mathop {\dot {f}}\nolimits_5 =559.1999942\mbox{Hzs}^{-1}$ listed in
Table 2, it is estimated that the chirp mass of the signal wave source is
$\mathscr{M}=36.090M_\odot $, and according to the lower frequency $f_6
=42.73112738\mbox{Hz}$ of the negative strain of the GW150914 signal wave
and its Lagrange change rate $\mathop {\dot {f}}\nolimits_6
=418.0919129\mbox{Hzs}^{-1}$, it is estimated that the chirp mass of the
signal wave source is $\mathscr{M}=30.873M_\odot $. It can be seen that the
estimations of chirp masses from different frequencies are very different,
and there is no basis for making trade-offs. This poses a challenge to the
general relativity Blanchet frequency equation. There are still unrecorded
lower frequencies of the signal wave, so there must be other different
estimates of the mass of the defect. Using the frequencies of all main
positive and negative strain peaks of the GW150914 signal wave and their
time derivatives, the estimated chirp masses of the wave source are
respectively as follows,
\end{multicols}
\begin{eqnarray*}
&& \mathscr{M}_2^+ =\frac{\left( {2.998\times 10^8} \right)^3}{6.674\times
10^{-11}}\left( {\frac{5}{96}\pi ^{-8/3}\times 45.376^{-11/3}\times 559.200}
\right)^{3/5}\frac{M_\odot }{1.989\times 10^{30}}=55.664M_\odot \\
&& \mathscr{M}_3^+ =\frac{\left( {2.998\times 10^8} \right)^3}{6.674\times
10^{-11}}\left( {\frac{5}{96}\pi ^{-8/3}\times 58.445^{-11/3}\times
1142.134} \right)^{3/5}\frac{M_\odot }{1.989\times 10^{30}}=48.960M_\odot \\
&& \mathscr{M}_4^+ =\frac{\left( {2.998\times 10^8} \right)^3}{6.674\times
10^{-11}}\left( {\frac{5}{96}\pi ^{-8/3}\times 79.318^{-11/3}\times
2747.764} \right)^{3/5}\frac{M_\odot }{1.989\times 10^{30}}=42.347M_\odot \\
&& \mathscr{M}_5^+ =\frac{\left( {2.998\times 10^8} \right)^3}{6.674\times
10^{-11}}\left( {\frac{5}{96}\pi ^{-8/3}\times 116.644^{-11/3}\times
8657.494} \right)^{3/5}\frac{M_\odot }{1.989\times 10^{30}}=36.090M_\odot \\
&& \mathscr{M}_2^ -  = \frac{{{{\left( {2.998 \times {{10}^8}} \right)}^3}}}{{6.674 \times {{10}^{ - 11}}}}{\left( {\frac{5}{{96}}{\pi ^{ - 8/3}} \times {{42.731}^{ - 11/3}} \times 418.092} \right)^{3/5}}\frac{{{M_ \odot }}}{{1.989 \times {{10}^{30}}}} = 53.356{M_ \odot } \hfill \\
&& \mathscr{M}_3^ -  = \frac{{{{\left( {2.998 \times {{10}^8}} \right)}^3}}}{{6.674 \times {{10}^{ - 11}}}}{\left( {\frac{5}{{96}}{\pi ^{ - 8/3}} \times {{53.045537}^{ - 11/3}} \times 764.196} \right)^{3/5}}\frac{{{M_ \odot }}}{{1.989 \times {{10}^{30}}}} = 47.616{M_ \odot } \hfill \\
&& \mathscr{M}_4^ -  = \frac{{{{\left( {2.998 \times {{10}^8}} \right)}^3}}}{{6.674 \times {{10}^{ - 11}}}}{\left( {\frac{5}{{96}}{\pi ^{ - 8/3}} \times {{68.323}^{ - 11/3}} \times 1560.827} \right)^{3/5}}\frac{{{M_ \odot }}}{{1.989 \times {{10}^{30}}}} = 41.881{M_ \odot } \hfill \\
&& \mathscr{M}_5^ -  = \frac{{{{\left( {2.998 \times {{10}^8}} \right)}^3}}}{{6.674 \times {{10}^{ - 11}}}}{\left( {\frac{5}{{96}}{\pi ^{ - 8/3}} \times {{92.724}^{ - 11/3}} \times 3755.062} \right)^{3/5}}\frac{{{M_ \odot }}}{{1.989 \times {{10}^{30}}}} = 36.224{M_ \odot } \hfill \\
&& \mathscr{M}_6^ -  = \frac{{{{\left( {2.998 \times {{10}^8}} \right)}^3}}}{{6.674 \times {{10}^{ - 11}}}}{\left( {\frac{5}{{96}}{\pi ^{ - 8/3}} \times {{136.358}^{ - 11/3}} \times 11831.23} \right)^{3/5}}\frac{{{M_ \odot }}}{{1.989 \times {{10}^{30}}}} = 30.873{M_ \odot } \hfill \\
\end{eqnarray*}
\begin{multicols}{2}

Among them, the positive and negative superscript indicates the calculation of positive and negative strains. The above approximate calculation results show that the frequency distribution of the positive and negative strain of the GW150914 signal wave determines two kinds of the chirp mass distribution of the wave source, which vary monotonously with the frequency of the change of the gravitational wave signal, and these values are very different. Before the merger, the rest mass of the two stars of the wave source of the GW150914 signal is invariable. The results of chirp mass estimation from different frequencies of the GW150914 signal wave are far from each other, and it is impossible to correct to be the approximate equivalent results. The results of numerical calculation above prove that there is a principled difference between the GW150914 signal wave and the gravitational wave theory of general relativity.

Since the estimation of the chirp mass is derived from the zero order approximation of the general relativistic Blanchet frequency equation, the above difference may be interpreted as an error caused by the zero order approximation. It seems that there is no such obvious difference between the original Blanchet frequency equation and the observed value, or there will be no obvious difference between the advanced approximation of the Blanchet frequency equation and the observed value. Is the result calculated according to the exact Blanchet frequency equation or its high-order approximation highly consistent with the GW150914 signal?

\section{Graphical proof of relativistic difference of GW150914 signal wave}

After detecting the gravitational wave signal and confirming that the wave source is a binary star gravitational system, the mass of the wave source can be determined by the frequency and strain distribution of the gravitational wave in theory. The Blanchet frequency equation is a highly nonlinear equation and cannot be solved accurately. However, in order to calculate the chirp mass, there is no scientific basis for choosing the zero order approximation of the equation under the low frequency condition. Ignoring the spin of the black hole, remove the spin-spin and spin-orbit interactions of the Blanchet frequency equation. The high-level approximation of the Blanchet equation is,
\end{multicols}
\begin{equation}
\label{eq5}
\begin{gathered}
  \pi \dot f = \frac{{96{G^{{5 \mathord{\left/
 {\vphantom {5 3}} \right.
 \kern-\nulldelimiterspace} 3}}}{m_1}{m_2}{{\left( {\pi f} \right)}^{{{11} \mathord{\left/
 {\vphantom {{11} 3}} \right.
 \kern-\nulldelimiterspace} 3}}}}}{{5{c^5}{{\left( {{m_1} + {m_2}} \right)}^{{1 \mathord{\left/
 {\vphantom {1 3}} \right.
 \kern-\nulldelimiterspace} 3}}}}} \times \left\{ {1 - \left[ {\frac{{743}}{{336}} + \frac{{11{m_1}{m_2}}}{{4{{\left( {{m_1} + {m_2}} \right)}^2}}}} \right]{{\left[ {\frac{{G\pi f\left( {{m_1} + {m_2}} \right)}}{{{c^3}}}} \right]}^{{2 \mathord{\left/
 {\vphantom {2 3}} \right.
 \kern-\nulldelimiterspace} 3}}}} \right. \hfill \\
  \quad \quad \left. { + 4\pi \left[ {\frac{{G\pi f}}{{{c^3}}}\left( {{m_1} + {m_2}} \right)} \right] + \left[ {\frac{{34103}}{{18144}} + \frac{{13661{m_1}{m_2}}}{{2016{{\left( {{m_1} + {m_2}} \right)}^2}}} + \frac{{59{{\left( {{m_1}{m_2}} \right)}^2}}}{{18{{\left( {{m_1} + {m_2}} \right)}^4}}}} \right]{{\left[ {\frac{{G\pi f}}{{{c^3}}}\left( {{m_1} + {m_2}} \right)} \right]}^{{4 \mathord{\left/
 {\vphantom {4 3}} \right.
 \kern-\nulldelimiterspace} 3}}}} \right\} \hfill \\
\end{gathered}
\end{equation}
\begin{multicols}{2}

Although this equation can not be solved accurately, the computer having developed to today, for some non-linear and implicit function equations which can not be solved accurately, approximate solutions with high accuracy can be obtained by numerical calculation or image solution to prove the reliability of qualitative conclusions. In this section, we study the difference between the frequency distribution of the GW150914 signal wave and the relativistic Blanchet frequency equation.

\subsection{Incompatibility between Blanchet curve and Lagrange polyline of GW150914 signal wave}

An image solution of the Blanchet equation (\ref{eq5}) is to give different mass of binary black holes in the equation, then draw the curves of the relation between frequency and frequency, and then draw the polyline of the relation between frequency and frequency of the GW150914 signal wave. If the latter has the tendency of coincidence or coincidence with one of the former, the mass of the wave source is determined. Otherwise, it is proved that the Blanchet frequency equation has no GW150914 signal wave solution.

As shown in Fig. 3, the polyline is the Lagrange frequency polyline (LC)
of the signal wave strain peak of GW150914. Seven sets of values
representing the mass of binary stars, $m_1 =29m_\odot ,\;m_2 =36m_\odot $;
$m_1 =7m_\odot ,\;m_2 =58m_\odot $; $m_1 =15m_\odot ,\;m_2 =50m_\odot $;
$m_1 =20m_\odot ,\;m_2 =45m_\odot $; $m_1 =10m_\odot ,\;m_2 =55m_\odot $;
$m_1 =10m_\odot ,\;m_2 =65m_\odot $; $m_1 =30m_\odot ,\;m_2 =50m_\odot $, in
which ${m_ \odot } = 1.989 \times {10^{30}}{\text{kg}}$ is the solar mass, are selected. The following seven
equations are obtained by substituting them into equation (5), respectively,

\end{multicols}
\small 
\begin{eqnarray*}
&&  \pi \dot f = \frac{{96{{\left( {6.674 \times {{10}^{ - 11}}} \right)}^{5/3}}\left( {29 \times 1.989 \times {{10}^{30}}} \right)\left( {36 \times 1.989 \times {{10}^{30}}} \right){{\left( {\pi  \times f} \right)}^{11/3}}}}{{5 \times {{\left( {2.998 \times {{10}^8}} \right)}^5}{{\left( {29 \times 1.989 \times {{10}^{30}} + 36 \times 1.989 \times {{10}^{30}}} \right)}^{1/3}}}} \hfill \\
&&  \quad \quad  \times \left\{ {1 - \left( {\frac{{743}}{{336}} + \frac{{11 \times 29 \times 36}}{{4{{\left( {29 + 36} \right)}^2}}}} \right){{\left( {\frac{{6.674 \times {{10}^{ - 11}}\pi  \times f \times \left( {29 + 36} \right) \times 1.989 \times {{10}^{30}}}}{{{{\left( {2.998 \times {{10}^8}} \right)}^3}}}} \right)}^{2/3}}} \right. \hfill \\
&&  \quad \quad  + 4\pi  \times \frac{{6.674 \times {{10}^{ - 11}}\pi  \times f \times \left( {29 + 36} \right) \times 1.989 \times {{10}^{30}}}}{{{{\left( {2.998 \times {{10}^8}} \right)}^3}}} \hfill \\
&&  \quad \quad \left. { + \left( {\frac{{34103}}{{18144}} + \frac{{13661 \times 29 \times 36}}{{2016{{\left( {29 + 36} \right)}^2}}} + \frac{{59 \times {{\left( {29 \times 36} \right)}^2}}}{{{{\left( {29 + 36} \right)}^4}}}} \right){{\left( {\frac{{6.674 \times {{10}^{ - 11}}\pi  \times f \times \left( {29 + 36} \right) \times 1.989 \times {{10}^{30}}}}{{{{\left( {2.998 \times {{10}^8}} \right)}^3}}}} \right)}^{4/3}}} \right\} \hfill \\
&&  \pi \dot f = \frac{{96{{\left( {6.674 \times {{10}^{ - 11}}} \right)}^{5/3}}\left( {7 \times 1.989 \times {{10}^{30}}} \right)\left( {58 \times 1.989 \times {{10}^{30}}} \right){{\left( {\pi  \times f} \right)}^{11/3}}}}{{5 \times {{\left( {2.998 \times {{10}^8}} \right)}^5}{{\left( {7 \times 1.989 \times {{10}^{30}} + 58 \times 1.989 \times {{10}^{30}}} \right)}^{1/3}}}} \hfill \\
&&  \quad \quad  \times \left\{ {1 - \left( {\frac{{743}}{{336}} + \frac{{11 \times 7 \times 58}}{{4{{\left( {7 + 58} \right)}^2}}}} \right){{\left( {\frac{{6.674 \times {{10}^{ - 11}}\pi  \times f \times \left( {7 + 58} \right) \times 1.989 \times {{10}^{30}}}}{{{{\left( {2.998 \times {{10}^8}} \right)}^3}}}} \right)}^{2/3}}} \right. \hfill \\
&&  \quad \quad  + 4\pi  \times \frac{{6.674 \times {{10}^{ - 11}}\pi  \times f \times \left( {7 + 58} \right) \times 1.989 \times {{10}^{30}}}}{{{{\left( {2.998 \times {{10}^8}} \right)}^3}}} \hfill \\
&&  \quad \quad \left. { + \left( {\frac{{34103}}{{18144}} + \frac{{13661 \times 7 \times 58}}{{2016{{\left( {7 + 58} \right)}^2}}} + \frac{{59 \times {{\left( {7 \times 58} \right)}^2}}}{{{{\left( {7 + 58} \right)}^4}}}} \right){{\left( {\frac{{6.674 \times {{10}^{ - 11}}\pi  \times f \times \left( {7 + 58} \right) \times 1.989 \times {{10}^{30}}}}{{{{\left( {2.998 \times {{10}^8}} \right)}^3}}}} \right)}^{4/3}}} \right\} \hfill \\
&&  \pi \dot f = \frac{{96{{\left( {6.674 \times {{10}^{ - 11}}} \right)}^{5/3}}\left( {15 \times 1.989 \times {{10}^{30}}} \right)\left( {50 \times 1.989 \times {{10}^{30}}} \right){{\left( {\pi  \times f} \right)}^{11/3}}}}{{5 \times {{\left( {2.998 \times {{10}^8}} \right)}^5}{{\left( {15 \times 1.989 \times {{10}^{30}} + 50 \times 1.989 \times {{10}^{30}}} \right)}^{1/3}}}} \hfill \\
&&  \quad \quad  \times \left\{ {1 - \left( {\frac{{743}}{{336}} + \frac{{11 \times 15 \times 50}}{{4{{\left( {15 + 50} \right)}^2}}}} \right){{\left( {\frac{{6.674 \times {{10}^{ - 11}}\pi  \times f \times \left( {15 + 50} \right) \times 1.989 \times {{10}^{30}}}}{{{{\left( {2.998 \times {{10}^8}} \right)}^3}}}} \right)}^{2/3}}} \right. \hfill \\
&&  \quad \quad  + 4\pi  \times \frac{{6.674 \times {{10}^{ - 11}}\pi  \times f \times \left( {15 + 50} \right) \times 1.989 \times {{10}^{30}}}}{{{{\left( {2.998 \times {{10}^8}} \right)}^3}}} \hfill \\
&&  \quad \quad \left. { + \left( {\frac{{34103}}{{18144}} + \frac{{13661 \times 15 \times 50}}{{2016{{\left( {15 + 50} \right)}^2}}} + \frac{{59 \times {{\left( {15 \times 50} \right)}^2}}}{{{{\left( {15 + 50} \right)}^4}}}} \right){{\left( {\frac{{6.674 \times {{10}^{ - 11}}\pi  \times f \times \left( {15 + 50} \right) \times 1.989 \times {{10}^{30}}}}{{{{\left( {2.998 \times {{10}^8}} \right)}^3}}}} \right)}^{4/3}}} \right\} \hfill \\
&&  \pi \dot f = \frac{{96{{\left( {6.674 \times {{10}^{ - 11}}} \right)}^{5/3}}\left( {25 \times 1.989 \times {{10}^{30}}} \right)\left( {40 \times 1.989 \times {{10}^{30}}} \right){{\left( {\pi  \times f} \right)}^{11/3}}}}{{5 \times {{\left( {2.998 \times {{10}^8}} \right)}^5}{{\left( {25 \times 1.989 \times {{10}^{30}} + 40 \times 1.989 \times {{10}^{30}}} \right)}^{1/3}}}} \hfill \\
&&  \quad \quad  \times \left\{ {1 - \left( {\frac{{743}}{{336}} + \frac{{11 \times 25 \times 40}}{{4{{\left( {25 + 40} \right)}^2}}}} \right){{\left( {\frac{{6.674 \times {{10}^{ - 11}}\pi  \times f \times \left( {25 + 40} \right) \times 1.989 \times {{10}^{30}}}}{{{{\left( {2.998 \times {{10}^8}} \right)}^3}}}} \right)}^{2/3}}} \right. \hfill \\
&&  \quad \quad  + 4\pi  \times \frac{{6.674 \times {{10}^{ - 11}}\pi  \times f \times \left( {25 + 40} \right) \times 1.989 \times {{10}^{30}}}}{{{{\left( {2.998 \times {{10}^8}} \right)}^3}}} \hfill \\
&&  \quad \quad \left. { + \left( {\frac{{34103}}{{18144}} + \frac{{13661 \times 25 \times 40}}{{2016{{\left( {25 + 40} \right)}^2}}} + \frac{{59 \times {{\left( {25 \times 40} \right)}^2}}}{{{{\left( {25 + 40} \right)}^4}}}} \right){{\left( {\frac{{6.674 \times {{10}^{ - 11}}\pi  \times f \times \left( {25 + 40} \right) \times 1.989 \times {{10}^{30}}}}{{{{\left( {2.998 \times {{10}^8}} \right)}^3}}}} \right)}^{4/3}}} \right\} \hfill \\
&&  \pi \dot f = \frac{{96{{\left( {6.674 \times {{10}^{ - 11}}} \right)}^{5/3}}\left( {10 \times 1.989 \times {{10}^{30}}} \right)\left( {55 \times 1.989 \times {{10}^{30}}} \right){{\left( {\pi  \times f} \right)}^{11/3}}}}{{5 \times {{\left( {2.998 \times {{10}^8}} \right)}^5}{{\left( {10 \times 1.989 \times {{10}^{30}} + 55 \times 1.989 \times {{10}^{30}}} \right)}^{1/3}}}} \hfill \\
&&  \quad \quad  \times \left\{ {1 - \left( {\frac{{743}}{{336}} + \frac{{11 \times 10 \times 55}}{{4{{\left( {10 + 55} \right)}^2}}}} \right){{\left( {\frac{{6.674 \times {{10}^{ - 11}}\pi  \times f \times \left( {10 + 55} \right) \times 1.989 \times {{10}^{30}}}}{{{{\left( {2.998 \times {{10}^8}} \right)}^3}}}} \right)}^{2/3}}} \right. \hfill \\
&&  \quad \quad  + 4\pi  \times \frac{{6.674 \times {{10}^{ - 11}}\pi  \times f \times \left( {10 + 55} \right) \times 1.989 \times {{10}^{30}}}}{{{{\left( {2.998 \times {{10}^8}} \right)}^3}}} \hfill \\
&&  \quad \quad \left. { + \left( {\frac{{34103}}{{18144}} + \frac{{13661 \times 10 \times 55}}{{2016{{\left( {10 + 55} \right)}^2}}} + \frac{{59 \times {{\left( {10 \times 55} \right)}^2}}}{{{{\left( {10 + 55} \right)}^4}}}} \right){{\left( {\frac{{6.674 \times {{10}^{ - 11}}\pi  \times f \times \left( {10 + 55} \right) \times 1.989 \times {{10}^{30}}}}{{{{\left( {2.998 \times {{10}^8}} \right)}^3}}}} \right)}^{4/3}}} \right\} \hfill \\
&&  \pi \dot f = \frac{{96{{\left( {6.674 \times {{10}^{ - 11}}} \right)}^{5/3}}\left( {10 \times 1.989 \times {{10}^{30}}} \right)\left( {65 \times 1.989 \times {{10}^{30}}} \right){{\left( {\pi  \times f} \right)}^{11/3}}}}{{5 \times {{\left( {2.998 \times {{10}^8}} \right)}^5}{{\left( {10 \times 1.989 \times {{10}^{30}} + 65 \times 1.989 \times {{10}^{30}}} \right)}^{1/3}}}} \hfill \\
&&  \quad \quad  \times \left\{ {1 - \left( {\frac{{743}}{{336}} + \frac{{11 \times 10 \times 65}}{{4{{\left( {10 + 65} \right)}^2}}}} \right){{\left( {\frac{{6.674 \times {{10}^{ - 11}}\pi  \times f \times \left( {10 + 65} \right) \times 1.989 \times {{10}^{30}}}}{{{{\left( {2.998 \times {{10}^8}} \right)}^3}}}} \right)}^{2/3}}} \right. \hfill \\
&&  \quad \quad  + 4\pi  \times \frac{{6.674 \times {{10}^{ - 11}}\pi  \times f \times \left( {10 + 65} \right) \times 1.989 \times {{10}^{30}}}}{{{{\left( {2.998 \times {{10}^8}} \right)}^3}}} \hfill \\
&&  \quad \quad \left. { + \left( {\frac{{34103}}{{18144}} + \frac{{13661 \times 10 \times 65}}{{2016{{\left( {10 + 65} \right)}^2}}} + \frac{{59 \times {{\left( {10 \times 65} \right)}^2}}}{{{{\left( {10 + 65} \right)}^4}}}} \right){{\left( {\frac{{6.674 \times {{10}^{ - 11}}\pi  \times f \times \left( {10 + 65} \right) \times 1.989 \times {{10}^{30}}}}{{{{\left( {2.998 \times {{10}^8}} \right)}^3}}}} \right)}^{4/3}}} \right\} \hfill \\
&&  \pi \dot f = \frac{{96{{\left( {6.674 \times {{10}^{ - 11}}} \right)}^{5/3}}\left( {30 \times 1.989 \times {{10}^{30}}} \right)\left( {50 \times 1.989 \times {{10}^{30}}} \right){{\left( {\pi  \times f} \right)}^{11/3}}}}{{5 \times {{\left( {2.998 \times {{10}^8}} \right)}^5}{{\left( {30 \times 1.989 \times {{10}^{30}} + 50 \times 1.989 \times {{10}^{30}}} \right)}^{1/3}}}} \hfill \\
&&  \quad \quad  \times \left\{ {1 - \left( {\frac{{743}}{{336}} + \frac{{11 \times 30 \times 50}}{{4{{\left( {30 + 50} \right)}^2}}}} \right){{\left( {\frac{{6.674 \times {{10}^{ - 11}}\pi  \times f \times \left( {30 + 50} \right) \times 1.989 \times {{10}^{30}}}}{{{{\left( {2.998 \times {{10}^8}} \right)}^3}}}} \right)}^{2/3}}} \right. \hfill \\
&&  \quad \quad  + 4\pi  \times \frac{{6.674 \times {{10}^{ - 11}}\pi  \times f \times \left( {30 + 50} \right) \times 1.989 \times {{10}^{30}}}}{{{{\left( {2.998 \times {{10}^8}} \right)}^3}}} \hfill \\
&&  \quad \quad \left. { + \left( {\frac{{34103}}{{18144}} + \frac{{13661 \times 30 \times 50}}{{2016{{\left( {30 + 50} \right)}^2}}} + \frac{{59 \times {{\left( {30 \times 50} \right)}^2}}}{{{{\left( {30 + 50} \right)}^4}}}} \right){{\left( {\frac{{6.674 \times {{10}^{ - 11}}\pi  \times f \times \left( {30 + 50} \right) \times 1.989 \times {{10}^{30}}}}{{{{\left( {2.998 \times {{10}^8}} \right)}^3}}}} \right)}^{4/3}}} \right\} \hfill \\
\end{eqnarray*}
\begin{multicols}{2}
\normalsize

These seven Blanchet frequency curves (BC) are also plotted in Figure 3.
Theoretically, the Lagrange frequency polyline of the signal wave should be
approximately coincident with the Blanchet frequency curve of the $36M_\odot
$and $29M_\odot $ mass combinations. Therefore, the masses of the two black
holes of the wave source can be found the best approximation from the curve.
However, the fact is that the Lagrange polylines of the GW150914 signal
waves deviate far from the Blanchet frequency curve of gravitational waves
from the binary black holes with masses of $36M_\odot $ and $29M_\odot $,
which is contradictory! More importantly, the Lagrange polyline of the
GW150914 signal wave cuts the Blanchet curves of different mass
combinations. This shows that there is no mass assignment of any group of
binary stars to make the Blanchet frequency curve coincide with the Lagrange
frequency polyline of the GW150914 signal wave. Therefore, the frequency
variation of the GW150914 signal wave does not satisfy the relativistic Blanchet
frequency equation at all. There is no scientific basis for anyone to be
able to claim that the GW150914 signal wave came from the merging process of
a pair binary black hole with 36 and 29 solar mass respectively.

Taking the combined values of different binary star masses $m_1 $ and $m_2
$ into the Blanche frequency equation can draw a lot of Blancchet frequency
curves. When the total mass is kept constant, the masses of the binary stars
are closer, and the Blanchet frequency curve is steeper with increasing
frequency. The trend of all these Blanchet curves and Lagrange frequency
polylines predicts that the Lagrange polyline must cut all the Blanchet
curves. It is also impossible to modify the
vibration curve of the GW150914 signal wave so that its frequency distribution
satisfies the Blanchet frequency equation. This is the general relativistic difference between the frequency distribution of the GW150914 signal wave and the Blanchet frequency equation of gravitational wave.
\begin{figure}[H]
\centerline{\includegraphics[width=\linewidth]{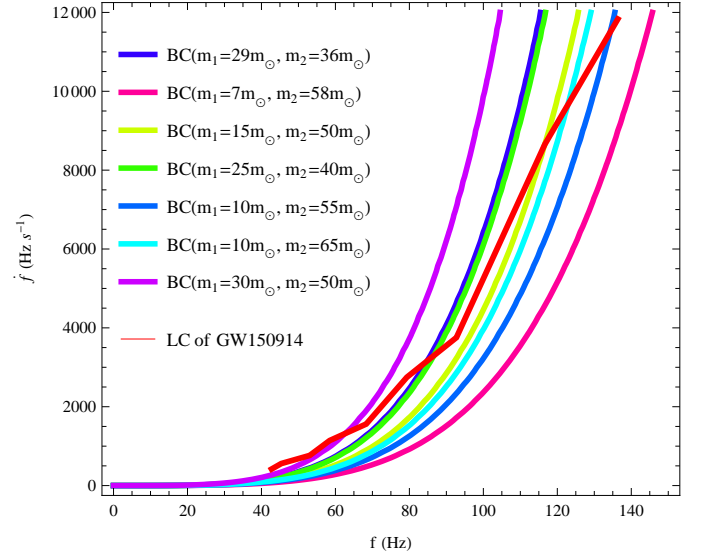}}
\caption{\footnotesize The Lagrange frequency polyline (LC) and the Blanchet frequency curve (BC) of the GW150914 signal wave. Graph lines intuitively reflect that the slope of each Blanchet frequency curve changes rapidly with increasing frequency, and the Lagrange frequency polyline cannot coincide with a certain Blancchet frequency curve, indicating that the GW150914 signal wave dos not support the generally relativistic Blanchet equation.}
\label{fig3}
\end{figure}

\subsection{Unsolvability of Blanchet equations for GW150914 signal wave}

The Lagrange frequency polyline of the GW150914 signal wave cutting the Blanchet
frequency curve of different mass combinations shows that the Blanchet
frequency equation does not have any solution that satisfies the GW150914
signal wave. This conclusion can also be rigorously proved by another
image solution.

Substituting the frequencies of the positive and negative strain peaks in
Table 1 and their Lagrange derivatives into the Blanche frequency equation
(\ref{eq5}) in order, a system of nine Blanchet equations is obtained,
\end{multicols}
\small 
\begin{eqnarray*} 
&&   \pi  \times 559.1999942 = \frac{{96{{\left( {6.674 \times {{10}^{ - 11}}} \right)}^{5/3}}\left( {{m_1} \times 1.989 \times {{10}^{30}}} \right)\left( {{m_2} \times 1.989 \times {{10}^{30}}} \right){{\left( {\pi  \times 45.37639637} \right)}^{11/3}}}}{{5 \times {{\left( {2.998 \times {{10}^8}} \right)}^5}{{\left( {{m_1} \times 1.989 \times {{10}^{30}} + {m_2} \times 1.989 \times {{10}^{30}}} \right)}^{1/3}}}} \hfill \\
&&  \quad \quad  \times \left\{ {1 - \left( {\frac{{743}}{{336}} + \frac{{11 {m_1} \times {m_2}}}{{4{{\left( {{m_1} + {m_2}} \right)}^2}}}} \right){{\left( {\frac{{6.674 \times {{10}^{ - 11}}\pi  \times 45.37639637\left( {{m_1} + {m_2}} \right) \times 1.989 \times {{10}^{30}}}}{{{{\left( {2.998 \times {{10}^8}} \right)}^3}}}} \right)}^{2/3}}} \right. \hfill \\
&&  \quad \quad  + 4\pi  \times \frac{{6.674 \times {{10}^{ - 11}}\pi  \times 45.37639637\left( {{m_1} + {m_2}} \right) \times 1.989 \times {{10}^{30}}}}{{{{\left( {2.998 \times {{10}^8}} \right)}^3}}} \hfill \\
&&    \left. {\quad \quad  + \left( {\frac{{34103}}{{18144}} + \frac{{13661{m_1}{m_2}}}{{2016{{\left( {{m_1} + {m_2}} \right)}^2}}} + \frac{{59 {{\left( {{m_1}{m_2}} \right)}^2}}}{{{{\left( {{m_1} + {m_2}} \right)}^4}}}} \right) \times {{\left( {\frac{{6.674 \times {{10}^{ - 11}}\pi  \times 45.37639637\left( {{m_1} + {m_2}} \right) \times 1.989 \times {{10}^{30}}}}{{{{\left( {2.998 \times {{10}^8}} \right)}^3}}}} \right)}^{4/3}}} \right\} \hfill \\
&&   \pi  \times 1142.134177 = \frac{{96{{\left( {6.674 \times {{10}^{ - 11}}} \right)}^{5/3}}\left( {{m_1} \times 1.989 \times {{10}^{30}}} \right)\left( {{m_2} \times 1.989 \times {{10}^{30}}} \right){{\left( {\pi  \times 58.44479852} \right)}^{11/3}}}}{{5 \times {{\left( {2.998 \times {{10}^8}} \right)}^5}{{\left( {{m_1} \times 1.989 \times {{10}^{30}} + {m_2} \times 1.989 \times {{10}^{30}}} \right)}^{1/3}}}} \hfill \\
&&  \quad \quad  \times \left\{ {1 - \left( {\frac{{743}}{{336}} + \frac{{11 {m_1}{m_2}}}{{4{{\left( {{m_1} + {m_2}} \right)}^2}}}} \right){{\left( {\frac{{6.674 \times {{10}^{ - 11}}\pi  \times 58.44479852\left( {{m_1} + {m_2}} \right) \times 1.989 \times {{10}^{30}}}}{{{{\left( {2.998 \times {{10}^8}} \right)}^3}}}} \right)}^{2/3}}} \right. \hfill \\
&&  \quad \quad  + 4\pi  \times \frac{{6.674 \times {{10}^{ - 11}}\pi  \times 58.44479852\left( {{m_1} + {m_2}} \right) \times 1.989 \times {{10}^{30}}}}{{{{\left( {2.998 \times {{10}^8}} \right)}^3}}} \hfill \\
&&  \quad \quad \left. { + \left( {\frac{{34103}}{{18144}} + \frac{{13661{m_1}{m_2}}}{{2016{{\left( {{m_1} + {m_2}} \right)}^2}}} + \frac{{59{{\left( {{m_1}{m_2}} \right)}^2}}}{{{{\left( {{m_1} + {m_2}} \right)}^4}}}} \right){{\left( {\frac{{6.674 \times {{10}^{ - 11}}\pi  \times 58.44479852\left( {{m_1} + {m_2}} \right) \times 1.989 \times {{10}^{30}}}}{{{{\left( {2.998 \times {{10}^8}} \right)}^3}}}} \right)}^{4/3}}} \right\} \hfill \\
&&   \pi  \times 2747.763841 = \frac{{96{{\left( {6.674 \times {{10}^{ - 11}}} \right)}^{5/3}}\left( {{m_1} \times 1.989 \times {{10}^{30}}} \right)\left( {{m_2} \times 1.989 \times {{10}^{30}}} \right){{\left( {\pi  \times 79.31794085} \right)}^{11/3}}}}{{5 \times {{\left( {2.998 \times {{10}^8}} \right)}^5}{{\left( {{m_1} \times 1.989 \times {{10}^{30}} + {m_2} \times 1.989 \times {{10}^{30}}} \right)}^{1/3}}}} \hfill \\
&&  \quad \quad  \times \left\{ {1 - \left( {\frac{{743}}{{336}} + \frac{{11 {m_1}{m_2}}}{{4{{\left( {{m_1} + {m_2}} \right)}^2}}}} \right){{\left( {\frac{{6.674 \times {{10}^{ - 11}}\pi  \times 79.31794085\left( {{m_1} + {m_2}} \right) \times 1.989 \times {{10}^{30}}}}{{{{\left( {2.998 \times {{10}^8}} \right)}^3}}}} \right)}^{2/3}}} \right. \hfill \\
&&  \quad \quad  + 4\pi  \times \frac{{6.674 \times {{10}^{ - 11}}\pi  \times 79.31794085\left( {{m_1} + {m_2}} \right) \times 1.989 \times {{10}^{30}}}}{{{{\left( {2.998 \times {{10}^8}} \right)}^3}}} \hfill \\
&&    \left. {\quad \quad  + \left( {\frac{{34103}}{{18144}} + \frac{{13661{m_1}{m_2}}}{{2016{{\left( {{m_1} + {m_2}} \right)}^2}}} + \frac{{59{{\left( {{m_1}{m_2}} \right)}^2}}}{{{{\left( {{m_1} + {m_2}} \right)}^4}}}} \right){{\left( {\frac{{6.674 \times {{10}^{ - 11}}\pi  \times 79.31794085\left( {{m_1} + {m_2}} \right) \times 1.989 \times {{10}^{30}}}}{{{{\left( {2.998 \times {{10}^8}} \right)}^3}}}} \right)}^{4/3}}} \right\} \hfill \\
&&   \pi  \times 8657.49422 = \frac{{96{{\left( {6.674 \times {{10}^{ - 11}}} \right)}^{5/3}}\left( {{m_1} \times 1.989 \times {{10}^{30}}} \right)\left( {{m_2} \times 1.989 \times {{10}^{30}}} \right){{\left( {\pi  \times 116.6440307} \right)}^{11/3}}}}{{5 \times {{\left( {2.998 \times {{10}^8}} \right)}^5}{{\left( {{m_1} \times 1.989 \times {{10}^{30}} + {m_2} \times 1.989 \times {{10}^{30}}} \right)}^{1/3}}}} \hfill \\
&&  \quad \quad  \times \left\{ {1 - \left( {\frac{{743}}{{336}} + \frac{{11 {m_1}{m_2}}}{{4{{\left( {{m_1} + {m_2}} \right)}^2}}}} \right){{\left( {\frac{{6.674 \times {{10}^{ - 11}}\pi  \times 116.6440307\left( {{m_1} + {m_2}} \right) \times 1.989 \times {{10}^{30}}}}{{{{\left( {2.998 \times {{10}^8}} \right)}^3}}}} \right)}^{2/3}}} \right. \hfill \\
&&  \quad \quad  + 4\pi  \times \frac{{6.674 \times {{10}^{ - 11}}\pi  \times 116.6440307\left( {{m_1} + {m_2}} \right) \times 1.989 \times {{10}^{30}}}}{{{{\left( {2.998 \times {{10}^8}} \right)}^3}}} \hfill \\
&&    \left. {\quad \quad  + \left( {\frac{{34103}}{{18144}} + \frac{{13661{m_1}{m_2}}}{{2016{{\left( {{m_1} + {m_2}} \right)}^2}}} + \frac{{59{{\left( {{m_1}{m_2}} \right)}^2}}}{{{{\left( {{m_1} + {m_2}} \right)}^4}}}} \right){{\left( {\frac{{6.674 \times {{10}^{ - 11}}\pi  \times 116.6440307\left( {{m_1} + {m_2}} \right) \times 1.989 \times {{10}^{30}}}}{{{{\left( {2.998 \times {{10}^8}} \right)}^3}}}} \right)}^{4/3}}} \right\} \hfill \\
&&   \pi  \times 418.0919129 = \frac{{96{{\left( {6.674 \times {{10}^{ - 11}}} \right)}^{5/3}}\left( {{m_1} \times 1.989 \times {{10}^{30}}} \right)\left( {{m_2} \times 1.989 \times {{10}^{30}}} \right){{\left( {\pi  \times 42.73112738} \right)}^{11/3}}}}{{5 \times {{\left( {2.998 \times {{10}^8}} \right)}^5}{{\left( {{m_1} \times 1.989 \times {{10}^{30}} + {m_2} \times 1.989 \times {{10}^{30}}} \right)}^{1/3}}}} \hfill \\
&&  \quad \quad  \times \left\{ {1 - \left( {\frac{{743}}{{336}} + \frac{{11 {m_1}{m_2}}}{{4{{\left( {{m_1} + {m_2}} \right)}^2}}}} \right){{\left( {\frac{{6.674 \times {{10}^{ - 11}}\pi  \times 42.73112738\left( {{m_1} + {m_2}} \right) \times 1.989 \times {{10}^{30}}}}{{{{\left( {2.998 \times {{10}^8}} \right)}^3}}}} \right)}^{2/3}}} \right. \hfill \\
&&  \quad \quad  + 4\pi  \times \frac{{6.674 \times {{10}^{ - 11}}\pi  \times 42.73112738\left( {{m_1} + {m_2}} \right) \times 1.989 \times {{10}^{30}}}}{{{{\left( {2.998 \times {{10}^8}} \right)}^3}}} \hfill \\
&&    \left. {\quad \quad  + \left( {\frac{{34103}}{{18144}} + \frac{{13661{m_1}{m_2}}}{{2016{{\left( {{m_1} + {m_2}} \right)}^2}}} + \frac{{59{{\left( {{m_1}{m_2}} \right)}^2}}}{{{{\left( {{m_1} + {m_2}} \right)}^4}}}} \right){{\left( {\frac{{6.674 \times {{10}^{ - 11}}\pi  \times 42.73112738\left( {{m_1} + {m_2}} \right) \times 1.989 \times {{10}^{30}}}}{{{{\left( {2.998 \times {{10}^8}} \right)}^3}}}} \right)}^{4/3}}} \right\} \hfill \\
&&   \pi  \times 764.1961764 = \frac{{96{{\left( {6.674 \times {{10}^{ - 11}}} \right)}^{5/3}}\left( {{m_1} \times 1.989 \times {{10}^{30}}} \right)\left( {{m_2} \times 1.989 \times {{10}^{30}}} \right){{\left( {\pi  \times 53.04553744} \right)}^{11/3}}}}{{5 \times {{\left( {2.998 \times {{10}^8}} \right)}^5}{{\left( {{m_1} \times 1.989 \times {{10}^{30}} + {m_2} \times 1.989 \times {{10}^{30}}} \right)}^{1/3}}}} \hfill \\
&&  \quad \quad  \times \left\{ {1 - \left( {\frac{{743}}{{336}} + \frac{{11 {m_1}{m_2}}}{{4{{\left( {{m_1} + {m_2}} \right)}^2}}}} \right){{\left( {\frac{{6.674 \times {{10}^{ - 11}}\pi  \times 53.04553744\left( {{m_1} + {m_2}} \right) \times 1.989 \times {{10}^{30}}}}{{{{\left( {2.998 \times {{10}^8}} \right)}^3}}}} \right)}^{2/3}}} \right. \hfill \\
&&  \quad \quad  + 4\pi  \times \frac{{6.674 \times {{10}^{ - 11}}\pi  \times 53.04553744\left( {{m_1} + {m_2}} \right) \times 1.989 \times {{10}^{30}}}}{{{{\left( {2.998 \times {{10}^8}} \right)}^3}}} \hfill \\
&&  \quad \quad \left. { + \left( {\frac{{34103}}{{18144}} + \frac{{13661{m_1}{m_2}}}{{2016{{\left( {{m_1} + {m_2}} \right)}^2}}} + \frac{{59{{\left( {{m_1}{m_2}} \right)}^2}}}{{{{\left( {{m_1} + {m_2}} \right)}^4}}}} \right){{\left( {\frac{{6.674 \times {{10}^{ - 11}}\pi  \times 53.04553744\left( {{m_1} + {m_2}} \right) \times 1.989 \times {{10}^{30}}}}{{{{\left( {2.998 \times {{10}^8}} \right)}^3}}}} \right)}^{4/3}}} \right\} \hfill \\
&&   \pi  \times 1560.827218 = \frac{{96{{\left( {6.674 \times {{10}^{ - 11}}} \right)}^{5/3}}\left( {{m_1} \times 1.989 \times {{10}^{30}}} \right)\left( {{m_2} \times 1.989 \times {{10}^{30}}} \right){{\left( {\pi  \times 68.32265222} \right)}^{11/3}}}}{{5 \times {{\left( {2.998 \times {{10}^8}} \right)}^5}{{\left( {{m_1} \times 1.989 \times {{10}^{30}} + {m_2} \times 1.989 \times {{10}^{30}}} \right)}^{1/3}}}} \hfill \\
&&  \quad \quad  \times \left\{ {1 - \left( {\frac{{743}}{{336}} + \frac{{11 {m_1}{m_2}}}{{4{{\left( {{m_1} + {m_2}} \right)}^2}}}} \right){{\left( {\frac{{6.674 \times {{10}^{ - 11}}\pi  \times 68.32265222\left( {{m_1} + {m_2}} \right) \times 1.989 \times {{10}^{30}}}}{{{{\left( {2.998 \times {{10}^8}} \right)}^3}}}} \right)}^{2/3}}} \right. \hfill \\
&&  \quad \quad  + 4\pi  \times \frac{{6.674 \times {{10}^{ - 11}}\pi  \times 68.32265222\left( {{m_1} + {m_2}} \right) \times 1.989 \times {{10}^{30}}}}{{{{\left( {2.998 \times {{10}^8}} \right)}^3}}} \hfill \\
&&  \quad \quad \left. { + \left( {\frac{{34103}}{{18144}} + \frac{{13661{m_1}{m_2}}}{{2016{{\left( {{m_1} + {m_2}} \right)}^2}}} + \frac{{59{{\left( {{m_1}{m_2}} \right)}^2}}}{{{{\left( {{m_1} + {m_2}} \right)}^4}}}} \right){{\left( {\frac{{6.674 \times {{10}^{ - 11}}\pi  \times 68.32265222\left( {{m_1} + {m_2}} \right) \times 1.989 \times {{10}^{30}}}}{{{{\left( {2.998 \times {{10}^8}} \right)}^3}}}} \right)}^{4/3}}} \right\} \hfill \\
&&   \pi  \times 3755.061951 = \frac{{96{{\left( {6.674 \times {{10}^{ - 11}}} \right)}^{5/3}}\left( {{m_1} \times 1.989 \times {{10}^{30}}} \right)\left( {{m_2} \times 1.989 \times {{10}^{30}}} \right){{\left( {\pi  \times 92.72359945} \right)}^{11/3}}}}{{5 \times {{\left( {2.998 \times {{10}^8}} \right)}^5}{{\left( {{m_1} \times 1.989 \times {{10}^{30}} + {m_2} \times 1.989 \times {{10}^{30}}} \right)}^{1/3}}}} \hfill \\
&&  \quad \quad  \times \left\{ {1 - \left( {\frac{{743}}{{336}} + \frac{{11 {m_1}{m_2}}}{{4{{\left( {{m_1} + {m_2}} \right)}^2}}}} \right){{\left( {\frac{{6.674 \times {{10}^{ - 11}}\pi  \times 92.72359945\left( {{m_1} + {m_2}} \right) \times 1.989 \times {{10}^{30}}}}{{{{\left( {2.998 \times {{10}^8}} \right)}^3}}}} \right)}^{2/3}}} \right. \hfill \\
&&  \quad \quad  + 4\pi  \times \frac{{6.674 \times {{10}^{ - 11}}\pi  \times 92.72359945\left( {{m_1} + {m_2}} \right) \times 1.989 \times {{10}^{30}}}}{{{{\left( {2.998 \times {{10}^8}} \right)}^3}}} \hfill \\
&&    \left. {\quad \quad  + \left( {\frac{{34103}}{{18144}} + \frac{{13661{m_1}{m_2}}}{{2016{{\left( {{m_1} + {m_2}} \right)}^2}}} + \frac{{59{{\left( {{m_1}{m_2}} \right)}^2}}}{{{{\left( {{m_1} + {m_2}} \right)}^4}}}} \right){{\left( {\frac{{6.674 \times {{10}^{ - 11}}\pi  \times 92.72359945\left( {{m_1} + {m_2}} \right) \times 1.989 \times {{10}^{30}}}}{{{{\left( {2.998 \times {{10}^8}} \right)}^3}}}} \right)}^{4/3}}} \right\} \hfill \\
&&   \pi  \times 11831.23042 = \frac{{96{{\left( {6.674 \times {{10}^{ - 11}}} \right)}^{5/3}}\left( {{m_1} \times 1.989 \times {{10}^{30}}} \right)\left( {{m_2} \times 1.989 \times {{10}^{30}}} \right){{\left( {\pi  \times 136.3582345} \right)}^{11/3}}}}{{5 \times {{\left( {2.998 \times {{10}^8}} \right)}^5}{{\left( {{m_1} \times 1.989 \times {{10}^{30}} + {m_2} \times 1.989 \times {{10}^{30}}} \right)}^{1/3}}}} \hfill \\
&&  \quad \quad  \times \left\{ {1 - \left( {\frac{{743}}{{336}} + \frac{{11 {m_1}{m_2}}}{{4{{\left( {{m_1} + {m_2}} \right)}^2}}}} \right){{\left( {\frac{{6.674 \times {{10}^{ - 11}}\pi  \times 136.3582345\left( {{m_1} + {m_2}} \right) \times 1.989 \times {{10}^{30}}}}{{{{\left( {2.998 \times {{10}^8}} \right)}^3}}}} \right)}^{2/3}}} \right. \hfill \\
&&  \quad \quad  + 4\pi  \times \frac{{6.674 \times {{10}^{ - 11}}\pi  \times 136.3582345\left( {{m_1} + {m_2}} \right) \times 1.989 \times {{10}^{30}}}}{{{{\left( {2.998 \times {{10}^8}} \right)}^3}}} \hfill \\
&&    \left. {\quad \quad  + \left( {\frac{{34103}}{{18144}} + \frac{{13661{m_1}{m_2}}}{{2016{{\left( {{m_1} + {m_2}} \right)}^2}}} + \frac{{59{{\left( {{m_1}{m_2}} \right)}^2}}}{{{{\left( {{m_1} + {m_2}} \right)}^4}}}} \right){{\left( {\frac{{6.674 \times {{10}^{ - 11}}\pi  \times 136.3582345\left( {{m_1} + {m_2}} \right) \times 1.989 \times {{10}^{30}}}}{{{{\left( {2.998 \times {{10}^8}} \right)}^3}}}} \right)}^{4/3}}} \right\} \hfill \\
\end{eqnarray*}
\begin{multicols}{2}
\normalsize
\begin{figure}[H]
\centerline{\includegraphics[width=\linewidth]{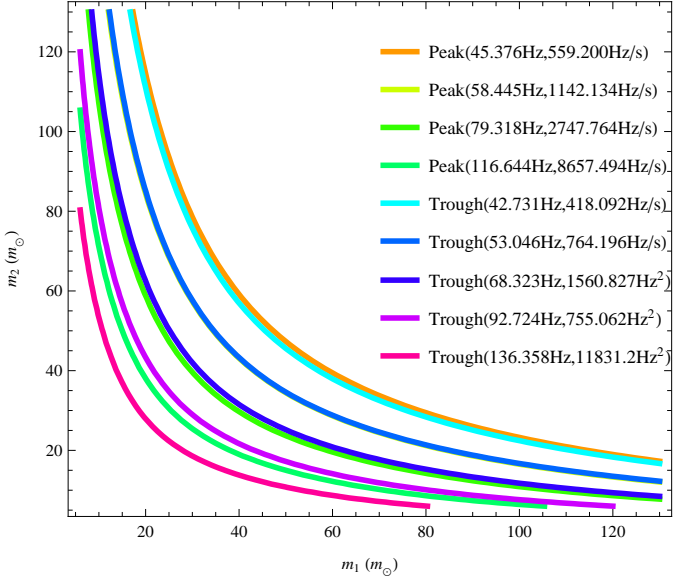}}
\caption{The Blanchet mass curves of the GW150914 signal wave. The group of
curves is similar to the isotherm without two intersection regions, showing
that the Blanchet frequency equation has no a GW150914 signal wave solution.}
\label{fig4}
\end{figure}

Theoretically, the solution of this system of equations must exist and be
unique. Otherwise, the result will prove that the Blanchet frequency
equation does not have a GW150914 signal wave solution. The unknowns of the
system of equations have only two masses, $m_1 $ and $m_2 $, and the
positions of the two can be exchanged. The curves of these equations are
plotted. According to the image solution, these curves should intersect in
two very small regions within the error range. The coordinates correspond to
the mass of the two black holes of the source. However, as shown in Fig. 4,
similar to the isotherm of the ideal gas in the closed container, the nine
Blanchet curves corresponding to the frequency and corresponding time
derivative of the signal wave strain peak of the GW150914 signal wave are disjoint, and the
results negate the existence and uniqueness of the solution to the equation,
that is, the GW150914 signal wave dos not support the Blanchet frequency
equation.

The image solution can explain intuitively and concisely whether the
specified signal wave solution of the non-linear Blanchet equation exists or
is unique. In fact, computer graphics can also give high-precision numerical
solutions. Therefore, the image solution is one of the best methods for
dealing with gravitational wave detection data and solving nonlinear Blancht
frequency equations. Observations of the GW150914 signal wave are
inconsistent with the results of the image solution of the Blanchet
frequency equation, which is a general relativity inference, and there is a
big difference between the general relativity prediction and the real
gravitational wave signal. This is the reason why a reasonable result cannot
be determined by using the Blanche frequency equation to estimate the chirp
mass of the GW150914 signal source.

In summary, if the GW150914 signal wave comes from the merger of two spiral
black holes with masses of 29 and 36 solar masses respectively, then the
Lagrange frequency polyline of the GW150914 signal wave will inevitably
coincide with the Blanchet frequency curve. However, the Lagrange frequency
polyline of the GW150914 signal wave cuts all the Blanchet frequency curves
of wave source mass combinations including 29 and 36 solar masses, so the
Blanchet frequency equation does not have a GW150914 signal wave solution.
However, the Lagrange frequency polyline of the GW150914 signal wave cuts
all the Blanchet frequency curves of different wave source mass combinations
including 29 and 36 solar masses, so the Blanchet frequency equation does
not have a GW150914 signal wave solution. The Blanchet frequency equation
set determined by the frequency and its time derivative contains only the
mass parameters, but the Blanch curve corresponding to the strain peak
frequency of the GW150914 signal wave is discrete without intersections,
which also indicates that the Blanchet frequency equation does not have
a GW150914 signal wave solution. There are indelible essential differences
between the GW150914 signal wave and the general relativistic Blanchet
frequency equation.

\section{Uncertainty of chirp mass of numerical relativistic waveform}

There is a difference between the frequency distribution law of the GW150914
signal wave and the general relativity Blanket frequency equation. Let us
now study the frequency distribution law of the so-called numerical
relativistic waveform of the GW150914 signal wave drawn by
LIGO\cite{} to understand the credibility of the numerical
relativistic waveform. As shown in Fig. 5, the time of the positive and
negative strain peaks of the numerical relativistic waveform of LIGO is
first extracted, and the corresponding periods, frequencies, and frequency
change rates are calculated. All the results are listed in Table 3. It is
shown from the calculation results that the $\dot {f}_i \dot {f}_i^{-2} $
values of the positive strain peaks and the negative strain peaks of the
numerical relativistic waveform do not have the same distribution, which
deviates from the law that the $\dot {f}_i \dot {f}_i^{-2} $ values of the
positive strain peaks and the negative strain peaks of the original GW150914
waveforms have the same distribution. On the other hand, the $\dot {f}_i \dot
{f}_i^{{-11} \mathord{\left/ {\vphantom {{-11} 3}} \right.
\kern-\nulldelimiterspace} 3} $ value of the numerical relativistic waveform
is also not a constant approximated by the zero-order approximation (\ref{eq3}) of
the Blanchet's frequency equation, and the value of $\dot {f}_i \dot
{f}_i^{{-11} \mathord{\left/ {\vphantom {{-11} 3}} \right.
\kern-\nulldelimiterspace} 3} $ at high frequencies is very different in
particular. It can be seen that the drawing of the numerical relativistic
waveform of GW150914 does not meet the requirement of logic self-consistent.
\begin{figure}[H]
\centerline{\includegraphics[width=\linewidth]{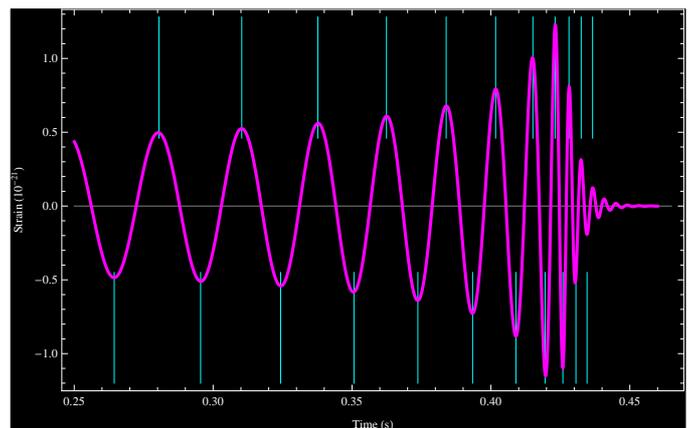}}
\caption{The time of the positive and negative strain peaks of the numerical
relativistic waveform of the GW150914 signal wave\cite{https:1}.}
\label{fig5}
\end{figure}

\end{multicols}
\begin{table}[H]
\renewcommand\arraystretch{0.9}
\begin{center}
\caption{The positive and negative strain frequencies and their
change rates of the numerical relativistic waveform}
\resizebox{1.0\textwidth}{!}{%
\begin{tabular}{c|ccccc|ccccc}
\toprule
\raisebox{-1.50ex}[0cm][0cm]{i}&
\multicolumn{5}{c|}{Positive strain of LIGO relativistic waveform} &
\multicolumn{5}{c}{Negative strain of LIGO relativistic waveform}  \\
\cline{2-11}
 &
$t_i $&
$f_i $&
$\dot {f}_i $&
${\dot {f}_i } \mathord{\left/ {\vphantom {{\dot {f}_i } {f_i^2 }}} \right. \kern-\nulldelimiterspace} {f_i^2 }$&
${\dot {f}_i } \mathord{\left/ {\vphantom {{\dot {f}_i } {f_i^{{11} \mathord{\left/ {\vphantom {{11} 3}} \right. \kern-\nulldelimiterspace} 3} }}} \right. \kern-\nulldelimiterspace} {f_i^{{11} \mathord{\left/ {\vphantom {{11} 3}} \right. \kern-\nulldelimiterspace} 3} }$&
$t_i $&
$f_i $&
$\dot {f}_i $&
${\dot {f}_i } \mathord{\left/ {\vphantom {{\dot {f}_i } {f_i^2 }}} \right. \kern-\nulldelimiterspace} {f_i^2 }$&
${\dot {f}_i } \mathord{\left/ {\vphantom {{\dot {f}_i } {f_i^{{11} \mathord{\left/ {\vphantom {{11} 3}} \right. \kern-\nulldelimiterspace} 3} }}} \right. \kern-\nulldelimiterspace} {f_i^{{11} \mathord{\left/ {\vphantom {{11} 3}} \right. \kern-\nulldelimiterspace} 3} }$ \\
\hline
1&
0.2805&
&
&
&
&
0.2644&
&
&
&
 \\
2&
0.3103&
33.55705&
&
&
&
0.2955&
&
&
&
 \\
3&
0.3377&
36.49635&
132.99008&
0.09984&
0.00024864&
0.3243&
32.15434&
&
&
 \\
4&
0.3624&
40.48583&
216.78090&
0.13226&
0.00027706&
0.3507&
34.72222&
103.70375&
0.08602&
0.00023276 \\
5&
0.3839&
46.51163&
399.33703&
0.18459&
0.00030687&
0.3737&
37.87879&
177.24775&
0.12353&
0.00028915 \\
6&
0.4017&
56.17978&
901.12942&
0.28551&
0.00034647&
0.3934&
43.47826&
301.70102&
0.15960&
0.00029688 \\
7&
0.4151&
74.62687&
3215.89835&
0.57745&
0.00043654&
0.409&
50.76142&
584.25788&
0.22674&
0.00032582 \\
8&
0.4231&
125.0000&
9644.08726&
0.61722&
0.00019751&
0.4195&
64.10256&
1704.08712&
0.41471&
0.00040391 \\
9&
0.4281&
200.0000&
10880.0774&
0.27200&
0.00003977&
0.4259&
95.23810&
5452.51100&
0.60114&
0.00030266 \\
10&
0.4325&
227.2727&
5164.99283&
0.09999&
0.00001181&
0.4306&
156.2500&
10588.0957&
0.43369&
0.00009568 \\
11&
0.4366&
243.9024&
&
&
&
0.4346&
212.7660&
10775.8621&
0.23804&
0.00003139
\\
\hline
\end{tabular}
}
\label{tab3}
\end{center}
\end{table}
\begin{multicols}{2}
\renewcommand\arraystretch{1.0}
\normalsize

The specific operational procedure of the numerical relativistic waveform
has not been disclosed, and the real physical meaning has not caused the
attention it deserves. The GW150914 signal wave was identified as coming
from a far-ancient spectacle of the merger of spiral binary black holes with
29$M_\odot $ and 36$M_\odot $ respectively, and mass of the combined black
hole is 62$M_\odot $. How to draw such a conclusion, the calculation process
of argument is missing. However, it is one of the key procedures to test the
theory of relativistic gravitational waves. Now we use the so-called
numerical relativistic gravitational waveform of the GW150914 signal wave to
estimate the chirp mass of the wave source. According to the values in Table
3, the peaks and troughs of the numerical relativistic waveform shown in
Figure 5 have eight Lagrange frequency derivatives, respectively, which are
substituted into the approximate equation (\ref{eq4}) to estimate the chirp mass.
The result has 16 different values,
\end{multicols}
\begin{eqnarray*}
&& \mathscr{M}_3^+ =\frac{\left( {2.998\times 10^8} \right)^3}{6.674\times
10^{-11}}\left( {\frac{5}{96}\pi ^{-8/3}\times 36.49635^{-11/3}\times
132.99008} \right)^{3/5}\frac{M_\odot }{1.989\times 10^{30}}=37.9661M_\odot
\\
&& \mathscr{M}_4^+ =\frac{\left( {2.998\times 10^8} \right)^3}{6.674\times
10^{-11}}\left( {\frac{5}{96}\pi ^{-8/3}\times 40.48583^{-11/3}\times
216.78090} \right)^{3/5}\frac{M_\odot }{1.989\times 10^{30}}=43.0746M_\odot
\\
&& \mathscr{M}_5^+ =\frac{\left( {2.998\times 10^8} \right)^3}{6.674\times
10^{-11}}\left( {\frac{5}{96}\pi ^{-8/3}\times 46.51163^{-11/3}\times
399.33703} \right)^{3/5}\frac{M_\odot }{1.989\times 10^{30}}=43.0746M_\odot
\\
&& \mathscr{M}_6^+ =\frac{\left( {2.998\times 10^8} \right)^3}{6.674\times
10^{-11}}\left( {\frac{5}{96}\pi ^{-8/3}\times 56.17978^{-11/3}\times
901.12942} \right)^{3/5}\frac{M_\odot }{1.989\times 10^{30}}=46.3286M_\odot
\\
&& \mathscr{M}_7^+ =\frac{\left( {2.998\times 10^8} \right)^3}{6.674\times
10^{-11}}\left( {\frac{5}{96}\pi ^{-8/3}\times 74.62687^{-11/3}\times
3215.89835} \right)^{3/5}\frac{M_\odot }{1.989\times 10^{30}}=53.2188M_\odot
\\
&& \mathscr{M}_8^+ =\frac{\left( {2.998\times 10^8} \right)^3}{6.674\times
10^{-11}}\left( {\frac{5}{96}\pi ^{-8/3}\times 125.00000^{-11/3}\times
9644.08726} \right)^{3/5}\frac{M_\odot }{1.989\times 10^{30}}=33.0680M_\odot
\\
&& \mathscr{M}_9^+ =\frac{\left( {2.998\times 10^8} \right)^3}{6.674\times
10^{-11}}\left( {\frac{5}{96}\pi ^{-8/3}\times 200.00000^{-11/3}\times
10880.07737} \right)^{3/5}\frac{M_\odot }{1.989\times
10^{30}}=12.6406M_\odot \\
&&  \mathscr{M}_{10}^+ =\frac{\left( {2.998\times 10^8} \right)^3}{6.674\times
10^{-11}}\left( {\frac{5}{96}\pi ^{-8/3}\times 227.27270^{-11/3}\times
5164.99283} \right)^{3/5}\frac{M_\odot }{1.989\times 10^{30}}=6.1023M_\odot
\\
&& \mathscr{M}_3^- =\frac{\left( {2.998\times 10^8} \right)^3}{6.674\times
10^{-11}}\left( {\frac{5}{96}\pi ^{-8/3}\times 34.72222^{-11/3}\times
103.70375} \right)^{3/5}\frac{M_\odot }{1.989\times 10^{30}}=36.4917M_\odot
\\
&& \mathscr{M}_4^- =\frac{\left( {2.998\times 10^8} \right)^3}{6.674\times
10^{-11}}\left( {\frac{5}{96}\pi ^{-8/3}\times 37.87879^{-11/3}\times
177.24775} \right)^{3/5}\frac{M_\odot }{1.989\times 10^{30}}=41.5652M_\odot
\\
&& \mathscr{M}_5^- =\frac{\left( {2.998\times 10^8} \right)^3}{6.674\times
10^{-11}}\left( {\frac{5}{96}\pi ^{-8/3}\times 43.47826^{-11/3}\times
301.70102} \right)^{3/5}\frac{M_\odot }{1.989\times 10^{30}}=42.2280M_\odot
\\
&& \mathscr{M}_6^- =\frac{\left( {2.998\times 10^8} \right)^3}{6.674\times
10^{-11}}\left( {\frac{5}{96}\pi ^{-8/3}\times 50.76142^{-11/3}\times
584.25788} \right)^{3/5}\frac{M_\odot }{1.989\times 10^{30}}=44.6521M_\odot
\\
&& \mathscr{M}_7^- =\frac{\left( {2.998\times 10^8} \right)^3}{6.674\times
10^{-11}}\left( {\frac{5}{96}\pi ^{-8/3}\times 64.10256^{-11/3}\times
1704.08712} \right)^{3/5}\frac{M_\odot }{1.989\times 10^{30}}=50.7952M_\odot
\\
&& \mathscr{M}_8^- =\frac{\left( {2.998\times 10^8} \right)^3}{6.674\times
10^{-11}}\left( {\frac{5}{96}\pi ^{-8/3}\times 95.23810^{-11/3}\times
5452.51100} \right)^{3/5}\frac{M_\odot }{1.989\times 10^{30}}=42.7197M_\odot
\\
&& \mathscr{M}_9^- =\frac{\left( {2.998\times 10^8} \right)^3}{6.674\times
10^{-11}}\left( {\frac{5}{96}\pi ^{-8/3}\times 156.25000^{-11/3}\times
10588.09569} \right)^{3/5}\frac{M_\odot }{1.989\times
10^{30}}=21.4062M_\odot \\
&& \mathscr{M}_{10}^- =\frac{\left( {2.998\times 10^8} \right)^3}{6.674\times
10^{-11}}\left( {\frac{5}{96}\pi ^{-8/3}\times 212.76600^{-11/3}\times
10775.86207} \right)^{3/5}\frac{M_\odot }{1.989\times
10^{30}}=10.9683M_\odot \\
\end{eqnarray*}
\begin{multicols}{2}
\normalsize

The estimate of the chirp mass corresponding to the numerical relativistic
waveform varies non-monotonically with frequency and cannot be approximately
equal, indicating that the frequency distribution of the numerical
relativistic waveform does not satisfy the general relativistic Blanche
equation. Although the concept of numerical relativistic wave has been
beautified and rendered, it has failed to achieve the expected goal.

LIGO announced the detection of gravitational waves in several spiral
binary black holes and one spiral binary neutron star system,
but avoided the calculation of the distribution and variation law of frequency,
which are the most basic physical quantities. The drawing procedure of numerical
relativistic gravitational waveform is also lack of necessary introduction.
All conclusions are only qualitative inferences. In fact, only after calculating
the frequency distribution and frequency change rate of the signal wave,
can we find out the law of frequency distribution and change, so as to determine
whether the detected signal wave really tests the general relativity theory.

\section{Classical estimation of total mass of GW150914 wave source}

The gravitational wave frequency of the same changing law can come from a
completely different wave source. The gravitational wave energy generated by
the combination of dense binary star can generate the spatiotemporal strain
with observation effect, causing the reaction of the detecting instrument.
The true gravitational wave signal detected by the precision detector plays
a decisive role in the development of gravitational theory. In fact, chirp
mass corresponds to the mass of countless pairs of stars. Estimating
chirp mass is not an effective way to understand wave source information,
especially when there are obvious contradictions in the estimation results.
Here we introduce the classical estimation results of the total mass of the wave source when the GW150914 signal wave is regarded as the gravitational wave of a system of spiral binary black hole.

As we all know, the calculation results of the classical mechanics of the
planetary revolution cycle are consistent with the astronomical
observations. A precise gravitational theory is bound to apply both to the
strong gravitational field and the weak gravitational field. As an
approximate analysis, the estimation of the total mass of a spiral binary
black hole by the classical equation of the orbit period will not be too far
from the predictions of any exact theory, otherwise there would be an exact
dividing point that does not actually exist between the strong gravitational
field and the weak gravitational field. If the binary black holes of mass
$m_1 $ and $m_2 $ are combined into one large black hole, it can be assumed
that a particle of mass $m_0 $ performs a uniform circular motion on the
Schwarzschild horizon\cite{Hawking:1972} of the large black hole. The
orbital frequency $f_s $ of the particle is the Schwarzschild frequency.
According to the gravitation Laws and circular motion laws are obtained,
\begin{equation}
\label{eq6}
\frac{G\left( {m_1 +m_2 } \right)m_0 }{R_s^2 }=m_0 R_s \left( {2\pi f_s }
\right)^2
\end{equation}

Among them, the Schwarzschild radius of a large black hole generated by a
binary black hole merger is $R_s ={2G\left( {m_1 +m_2 } \right)}
\mathord{\left/ {\vphantom {{2G\left( {m_1 +m_2 } \right)} {c^2}}} \right.
\kern-\nulldelimiterspace} {c^2}$, which is replaced by the above formula to
get
\begin{equation}
\label{eq7}
m_1 +m_2 =\frac{c^3}{2^{5 \mathord{\left/ {\vphantom {5 2}} \right.
\kern-\nulldelimiterspace} 2}\pi Gf_s }
\end{equation}

The gravitational radiation generated by the microscopic particles around
the dense star is very small and has no observation effect. Therefore, it is
usually assumed that two spiral black holes merge to generate strong
gravitational radiation. The Schwarzschild radii of the two black holes
before the merger are $R_{s1} ={2Gm_1 } \mathord{\left/ {\vphantom {{2Gm_1 }
{c^2}}} \right. \kern-\nulldelimiterspace} {c^2}$ and $R_{s2} ={2Gm_2 }
\mathord{\left/ {\vphantom {{2Gm_2 } {c^2}}} \right.
\kern-\nulldelimiterspace} {c^2}$, respectively. The two black holes run
around their mass centers, and detoured Schwarzschild frequency $f_s $ still
satisfies equation (\ref{eq7}).

Assuming that the maximum frequency of the GW150914 signal wave is the
Schwarzschild frequency $f_s $ of the wave source, then substituting the
maximum frequency $f_s =230.760\mbox{Hz}$ of the negative strain in Table 1
into equation (\ref{eq7}) gives the total mass $m_1 +m_2 =9.845\times
10^{31}\mbox{kg}=49.497M_\odot $ of two black holes or binary compact stars,
where the mass of the sun is $M_\odot =1.989\times 10^{30}\mbox{kg}$. And
the other estimate of the total mass of the two black holes obtained by
substituting the maximum frequency $f_s =197.398\mbox{Hz}$ of the positive
strain in Table 1 into equation (\ref{eq7}) is $59.43M_\odot $. There are some
differences between the two estimates. We choose the maximum value that
meets the expectations,
\begin{equation}
\label{eq8}
m_1 +m_2 =59.43M_\odot
\end{equation}

The chirp mass of the source estimated by Blanchet frequency equation is very uncertain, so the uncertainty of the total mass is also very large. But the uncertainty of estimating the total mass of the binary black hole according to the Schwarzschild orbit frequency is relatively small. Of course this is just an estimate. Accurate calculation belongs to the content of com quantum theory, which systematically studies the laws of com quantization in nature, what is given therein is the unique value of the statistical average.

\section{Conclusions and comments}

Scientific problems must ultimately be solved through scientific calculation. The important feature of GW150914 signal is the monotonic increase in frequency. We studied in detail the relationship between the frequency distribution of GW150914 signal wave and the generalized relativistic Blanchet frequency equation. It was pointed out that the similarity between GW150914 signal wave and the wave predicted by general relativity is only qualitative. However, frequency distribution and variation law of GW150914 signal do not support the non-linear Blanchet frequency equation, and the difference between them is far beyond the error range. On the other hand, the numerical relativistic waveform deviates too far from the original GW150914 signal waveform.

There is no precise demarcation point between the strong gravitational field and the weak gravitational field. The classical theory of gravitation is based on a large number of astronomical observations. There is no principle difference between the inferences of classical theory describing the gravitational system of black holes and the correct inference beyond classical theory. Therefore, if GW150914 signal wave belongs to gravitational wave of spiral binary stars, the Schwarzschild orbital frequency can be combined with classical theory to estimate the total mass of the wave source. The problem also restores its simple and easy-to-understand nature. The maximum frequency of positive strain of GW150914 signal wave is used to estimate the total mass of the wave source. The result is in line with expectation, but the maximum frequency of negative strain of GW150914 is used to estimate the total mass of the wave source, the result is not in line with expectation. Is there any scientific basis for making a unique choice between the two estimation of the total mass of the wave source? What kind of exact equation does the frequency of the GW150914 signal wave satisfy? How to accurately calculate the gravitational wave source mass and determine the exact position of the gravitational wave source? How to accurately distinguish the different gravitational wave signals from the binary black hole, the dense binary star, the multi black hole or the dense multi star gravitational system? What are the necessary and sufficient conditions leading to the formation and merging of spiral binary black holes? All these are urgent problems to be solved by gravitational theory.

Although the frequency distribution of the GW150914 signal wave accords with the motion law of the classical process of spiral binary star, the numerical calculation results of the discrete frequency and rate of change of the positive and negative strain show that only use the quantum number can accurately describe the law of gravitational wave. This is the com quantum theory which is different from the traditional quantum theory. An accurate theory of gravitational waves is bound to be highly consistent with the exact results of experimental observations. The detection of GW150914 signal wave is a scientific fortune, and this achievement will play an important role in the history of gravitational theory for a long time. Based on the monotonic increase of the frequency and strain of the signal wave, it can only be qualitatively judged that the signal wave detected by the laser interferometer gravitational wave detector belongs to the gravitational wave combined by a spiral binary black hole or a helical binary neutron star. The strict proof of the conclusion requires that the numerical analysis results of the observation data with high accuracy conform to the theoretical equation. Recognition of frequency-varying signal wave is a new technology to be developed. The generalized quantization characteristics of GW150914 signal contain a new scientific theory. Perhaps gravitational waves of spiral binaries that will be discovered in the future have the same general quantization law, and the gravitation theory will be developed and perfected.


\end{multicols}

\centerline{\rule{90mm}{0.8pt}}

\begin{multicols}{2}

\end{multicols}

\begin{appendix}
\section{Mathematica Code of FIG3}
\noindent\(\pmb{\text{}}\\
{\text{Show}\left[ \text{ContourPlot}\left[\left\{\pi \dot{f}==\frac{96\left(6.674\times 10^{-11}\right)^{5/3}\left(29\times 1.989\times 10^{30}\right)\left(36\times
1.989\times 10^{30}\right)(\pi \times f)^{11/3}}{5\times \left(2.998\times 10^8\right)^5\left(29\times 1.989\times 10^{30}+36\times 1.989\times 10^{30}\right)^{1/3}}\right.\right.\right.}\\
{\times \left\{1-\left(\frac{743}{336}+\frac{11\times 29\times 36}{4(29+36)^2}\right)\left(\frac{6.674\times 10^{-11}\pi \times f\times (29+36)\times
1.989\times 10^{30}}{\left(2.998\times 10^8\right)^3}\right)^{2/3}+4\pi \times \frac{6.674\times 10^{-11}\pi \times f\times (29+36)\times 1.989\times
10^{30}}{\left(2.998\times 10^8\right)^3}\right.}\\
{\left.+\left(\frac{34103}{18144}+\frac{13661\times 29\times 36}{2016(29+36)^2}+\frac{59\times (29\times 36)^2}{(29+36)^4}\right)\left(\frac{6.674\times
10^{-11}\pi \times f\times (29+36)\times 1.989\times 10^{30}}{\left(2.998\times 10^8\right)^3}\right)^{4/3}\right\},}\\
{\pi \dot{f}==\frac{96\left(6.674\times 10^{-11}\right)^{5/3}\left(7\times 1.989\times 10^{30}\right)\left(58\times 1.989\times 10^{30}\right)(\pi
\times f)^{11/3}}{5\times \left(2.998\times 10^8\right)^5\left(7\times 1.989\times 10^{30}+58\times 1.989\times 10^{30}\right)^{1/3}}}\\
{\times \left\{1-\left(\frac{743}{336}+\frac{11\times 7\times 58}{4(7+58)^2}\right)\left(\frac{6.674\times 10^{-11}\pi \times f\times (7+58)\times
1.989\times 10^{30}}{\left(2.998\times 10^8\right)^3}\right)^{2/3}+4\pi \times \frac{6.674\times 10^{-11}\pi \times f\times (7+58)\times 1.989\times
10^{30}}{\left(2.998\times 10^8\right)^3}\right.}\\
{\left.+\left(\frac{34103}{18144}+\frac{13661\times 7\times 58}{2016(7+58)^2}+\frac{59\times (7\times 58)^2}{(7+58)^4}\right)\left(\frac{6.674\times
10^{-11}\pi \times f\times (7+58)\times 1.989\times 10^{30}}{\left(2.998\times 10^8\right)^3}\right)^{4/3}\right\},}\\
{\pi \dot{f}==\frac{96\left(6.674\times 10^{-11}\right)^{5/3}\left(15\times 1.989\times 10^{30}\right)\left(50\times 1.989\times 10^{30}\right)(\pi
\times f)^{11/3}}{5\times \left(2.998\times 10^8\right)^5\left(15\times 1.989\times 10^{30}+50\times 1.989\times 10^{30}\right)^{1/3}}}\\
{\times \left\{1-\left(\frac{743}{336}+\frac{11\times 15\times 50}{4(15+50)^2}\right)\left(\frac{6.674\times 10^{-11}\pi \times f\times (15+50)\times
1.989\times 10^{30}}{\left(2.998\times 10^8\right)^3}\right)^{2/3}+4\pi \times \frac{6.674\times 10^{-11}\pi \times f\times (15+50)\times 1.989\times
10^{30}}{\left(2.998\times 10^8\right)^3}\right.}\\
{\left.+\left(\frac{34103}{18144}+\frac{13661\times 15\times 50}{2016(15+50)^2}+\frac{59\times (15\times 50)^2}{(15+50)^4}\right)\left(\frac{6.674\times
10^{-11}\pi \times f\times (15+50)\times 1.989\times 10^{30}}{\left(2.998\times 10^8\right)^3}\right)^{4/3}\right\},}\\
{\pi \dot{f}==\frac{96\left(6.674\times 10^{-11}\right)^{5/3}\left(25\times 1.989\times 10^{30}\right)\left(40\times 1.989\times 10^{30}\right)(\pi
\times f)^{11/3}}{5\times \left(2.998\times 10^8\right)^5\left(25\times 1.989\times 10^{30}+40\times 1.989\times 10^{30}\right)^{1/3}}}\\
{\times \left\{1-\left(\frac{743}{336}+\frac{11\times 25\times 40}{4(25+40)^2}\right)\left(\frac{6.674\times 10^{-11}\pi \times f\times (25+40)\times
1.989\times 10^{30}}{\left(2.998\times 10^8\right)^3}\right)^{2/3}+4\pi \times \frac{6.674\times 10^{-11}\pi \times f\times (25+40)\times 1.989\times
10^{30}}{\left(2.998\times 10^8\right)^3}\right.}\\
{\left.+\left(\frac{34103}{18144}+\frac{13661\times 25\times 40}{2016(25+40)^2}+\frac{59\times (25\times 40)^2}{(25+40)^4}\right)\left(\frac{6.674\times
10^{-11}\pi \times f\times (25+40)\times 1.989\times 10^{30}}{\left(2.998\times 10^8\right)^3}\right)^{4/3}\right\},}\\
{\pi \dot{f}==\frac{96\left(6.674\times 10^{-11}\right)^{5/3}\left(10\times 1.989\times 10^{30}\right)\left(55\times 1.989\times 10^{30}\right)(\pi
\times f)^{11/3}}{5\times \left(2.998\times 10^8\right)^5\left(10\times 1.989\times 10^{30}+55\times 1.989\times 10^{30}\right)^{1/3}}}\\
{\times \left\{1-\left(\frac{743}{336}+\frac{11\times 10\times 55}{4(10+55)^2}\right)\left(\frac{6.674\times 10^{-11}\pi \times f\times (10+55)\times
1.989\times 10^{30}}{\left(2.998\times 10^8\right)^3}\right)^{2/3}+4\pi \times \frac{6.674\times 10^{-11}\pi \times f\times (10+55)\times 1.989\times
10^{30}}{\left(2.998\times 10^8\right)^3}\right.}\\
{\left.+\left(\frac{34103}{18144}+\frac{13661\times 10\times 55}{2016(10+55)^2}+\frac{59\times (10\times 55)^2}{(10+55)^4}\right)\left(\frac{6.674\times
10^{-11}\pi \times f\times (10+55)\times 1.989\times 10^{30}}{\left(2.998\times 10^8\right)^3}\right)^{4/3}\right\},}\\
{\pi \dot{f}==\frac{96\left(6.674\times 10^{-11}\right)^{5/3}\left(10\times 1.989\times 10^{30}\right)\left(65\times 1.989\times 10^{30}\right)(\pi
\times f)^{11/3}}{5\times \left(2.998\times 10^8\right)^5\left(10\times 1.989\times 10^{30}+65\times 1.989\times 10^{30}\right)^{1/3}}}\\
{\times \left\{1-\left(\frac{743}{336}+\frac{11\times 10\times 65}{4(10+65)^2}\right)\left(\frac{6.674\times 10^{-11}\pi \times f\times (10+65)\times
1.989\times 10^{30}}{\left(2.998\times 10^8\right)^3}\right)^{2/3}+4\pi \times \frac{6.674\times 10^{-11}\pi \times f\times (10+65)\times 1.989\times
10^{30}}{\left(2.998\times 10^8\right)^3}\right.}\\
{\left.+\left(\frac{34103}{18144}+\frac{13661\times 10\times 65}{2016(10+65)^2}+\frac{59\times (10\times 65)^2}{(10+65)^4}\right)\left(\frac{6.674\times
10^{-11}\pi \times f\times (10+65)\times 1.989\times 10^{30}}{\left(2.998\times 10^8\right)^3}\right)^{4/3}\right\},}\\
{\pi \dot{f}==\frac{96\left(6.674\times 10^{-11}\right)^{5/3}\left(30\times 1.989\times 10^{30}\right)\left(50\times 1.989\times 10^{30}\right)(\pi
\times f)^{11/3}}{5\times \left(2.998\times 10^8\right)^5\left(30\times 1.989\times 10^{30}+50\times 1.989\times 10^{30}\right)^{1/3}}}\\
{\times \left\{1-\left(\frac{743}{336}+\frac{11\times 30\times 50}{4(30+50)^2}\right)\left(\frac{6.674\times 10^{-11}\pi \times f\times (30+50)\times
1.989\times 10^{30}}{\left(2.998\times 10^8\right)^3}\right)^{2/3}+4\pi \times \frac{6.674\times 10^{-11}\pi \times f\times (30+50)\times 1.989\times
10^{30}}{\left(2.998\times 10^8\right)^3}\right.}\\
{\left.\left.+\left(\frac{34103}{18144}+\frac{13661\times 30\times 50}{2016(30+50)^2}+\frac{59\times (30\times 50)^2}{(30+50)^4}\right)\left(\frac{6.674\times
10^{-11}\pi \times f\times (30+50)\times 1.989\times 10^{30}}{\left(2.998\times 10^8\right)^3}\right)^{4/3}\right\}\right\},}\\
{\{f,0,150\},\left\{\dot{f},0,12000\right\},\text{Axes}\to \text{True},\text{GridLinesStyle}\to \text{Directive}[\text{Dashed}],}\\
{\text{PlotLegends}\to \text{Placed}\left[\left\{\text{$\texttt{"}$BC(}m_1\text{=29}m_{\odot }\text{, }m_2\text{=36}m_{\odot }\text{)$\texttt{"}$},
\text{$\texttt{"}$BC(}m_1\text{=7}m_{\odot }\text{, }m_2\text{=58}m_{\odot }\text{)$\texttt{"}$},\right.\right.}\\
{\text{$\texttt{"}$BC(}m_1\text{=15}m_{\odot }\text{, }m_2\text{=50}m_{\odot }\text{)$\texttt{"}$}, \text{$\texttt{"}$BC(}m_1\text{=25}m_{\odot
}\text{, }m_2\text{=40}m_{\odot }\text{)$\texttt{"}$}, \text{$\texttt{"}$BC(}m_1\text{=10}m_{\odot }\text{, }m_2\text{=55}m_{\odot }\text{)$\texttt{"}$},}\\
{\left.\left.\text{$\texttt{"}$BC(}m_1\text{=10}m_{\odot }\text{, }m_2\text{=65}m_{\odot }\text{)$\texttt{"}$},\text{$\texttt{"}$BC(}m_1\text{=30}m_{\odot
}\text{, }m_2\text{=50}m_{\odot }\text{)$\texttt{"}$}\right\},\{0.250,0.60\}\right],\text{Ticks}\to \{\text{Range}[-10,10,2]\},}\\
{\text{ContourStyle}\to \{\{\text{Hue}[0.7],\text{AbsoluteThickness}[3]\},\{\text{Hue}[0.9],\text{AbsoluteThickness}[3]\},}\\
{\{\text{Hue}[0.2],\text{AbsoluteThickness}[3]\},\{\text{Hue}[0.3],\text{AbsoluteThickness}[3]\},}\\
{\{\text{Hue}[0.6],\text{AbsoluteThickness}[3]\},\{\text{Hue}[0.5],\text{AbsoluteThickness}[3]\},\{\text{Hue}[0.8],\text{AbsoluteThickness}[3]\}\},}\\
{\left.\text{FrameLabel}\to \left\{\text{{``}f (Hz){''}},\texttt{"}\dot{f}\text{ (Hz }s^{-1}\text{)$\texttt{"}$}\right\}\right],}\\
{\text{ListLinePlot}[\{\{136.3582345,11831.23042\},\{116.6440307,8657.49422\},}\\
{\{92.72359945,3755.061951\},\{79.31794085,2747.763841\},\{68.32265222,1560.827218\},}\\
{\{58.44479852,1142.134177\},\{53.04553744,764.1961764\},\{45.37639637,559.1999942\}, }\\
{\{42.73112738,418.0919129\}\},\text{PlotStyle}\to \{\text{Thickness}[0.010],\text{RGBColor}[1,0,0]\},}\\
{\text{PlotLegends}\to \text{Placed}[\{\text{{``}LC of GW150914{''}}\},\{0.250,0.60\}]],\text{AspectRatio}\to 0.86]}\)

\section{Mathematica Code of FIG4}
\noindent\({\text{ContourPlot}\left[\left\{\pi \times 559.1999942==\frac{96\left(6.674\times 10^{-11}\right)^{5/3}\left(m_1\times 1.989\times
10^{30}\right)\left(m_2\times 1.989\times 10^{30}\right)(\pi \times 45.37639637)^{11/3}}{5\times \left(2.998\times 10^8\right)^5\left(m_1 \times
1.989\times 10^{30}+m_2\times 1.989\times 10^{30}\right){}^{1/3}}\right.\right.}\\
{\times \left\{1-\left(\frac{743}{336}+\frac{11\times m_1\times m_2}{4\left(m_1+m_2\right){}^2}\right)\left(\frac{6.674\times 10^{-11}\pi \times
45.37639637\left(m_1+m_2\right)\times 1.989\times 10^{30}}{\left(2.998\times 10^8\right)^3}\right){}^{2/3}\right.}\\
{+4\pi \times \frac{6.674\times 10^{-11}\pi \times 45.37639637\left(m_1+m_2\right)\times 1.989\times 10^{30}}{\left(2.998\times 10^8\right)^3}}\\
{\left.+\left(\frac{34103}{18144}+\frac{13661\times m_1\times m_2}{2016\left(m_1+m_2\right){}^2}+\frac{59\times \left(m_1\times m_2\right){}^2}{\left(m_1+m_2\right){}^4}\right)\left(\frac{6.674\times
10^{-11}\pi \times 45.37639637\left(m_1+m_2\right)\times 1.989\times 10^{30}}{\left(2.998\times 10^8\right)^3}\right){}^{4/3}\right\},}\\
{\pi \times 1142.134177==\frac{96\left(6.674\times 10^{-11}\right)^{5/3}\left(m_1\times 1.989\times 10^{30}\right)\left(m_2\times 1.989\times
10^{30}\right)(\pi \times 58.44479852)^{11/3}}{5\times \left(2.998\times 10^8\right)^5\left(m_1 \times 1.989\times 10^{30}+m_2\times 1.989\times
10^{30}\right){}^{1/3}}}\\
{\times \left\{1-\left(\frac{743}{336}+\frac{11\times m_1\times m_2}{4\left(m_1+m_2\right){}^2}\right)\left(\frac{6.674\times 10^{-11}\pi \times
58.44479852\left(m_1+m_2\right)\times 1.989\times 10^{30}}{\left(2.998\times 10^8\right)^3}\right){}^{2/3}\right.}\\
{+4\pi \times \frac{6.674\times 10^{-11}\pi \times 58.44479852\left(m_1+m_2\right)\times 1.989\times 10^{30}}{\left(2.998\times 10^8\right)^3}}\\
{\left.+\left(\frac{34103}{18144}+\frac{13661\times m_1\times m_2}{2016\left(m_1+m_2\right){}^2}+\frac{59\times \left(m_1\times m_2\right){}^2}{\left(m_1+m_2\right){}^4}\right)\left(\frac{6.674\times
10^{-11}\pi \times 58.44479852\left(m_1+m_2\right)\times 1.989\times 10^{30}}{\left(2.998\times 10^8\right)^3}\right){}^{4/3}\right\},}\\
{\pi \times 2747.763841==\frac{96\left(6.674\times 10^{-11}\right)^{5/3}\left(m_1\times 1.989\times 10^{30}\right)\left(m_2\times 1.989\times
10^{30}\right)(\pi \times 79.31794085)^{11/3}}{5\times \left(2.998\times 10^8\right)^5\left(m_1 \times 1.989\times 10^{30}+m_2\times 1.989\times
10^{30}\right){}^{1/3}}}\\
{\times \left\{1-\left(\frac{743}{336}+\frac{11\times m_1\times m_2}{4\left(m_1+m_2\right){}^2}\right)\left(\frac{6.674\times 10^{-11}\pi \times
79.31794085\left(m_1+m_2\right)\times 1.989\times 10^{30}}{\left(2.998\times 10^8\right)^3}\right){}^{2/3}\right.}\\
{+4\pi \times \frac{6.674\times 10^{-11}\pi \times 79.31794085\left(m_1+m_2\right)\times 1.989\times 10^{30}}{\left(2.998\times 10^8\right)^3}}\\
{\left.+\left(\frac{34103}{18144}+\frac{13661\times m_1\times m_2}{2016\left(m_1+m_2\right){}^2}+\frac{59\times \left(m_1\times m_2\right){}^2}{\left(m_1+m_2\right){}^4}\right)\left(\frac{6.674\times
10^{-11}\pi \times 79.31794085\left(m_1+m_2\right)\times 1.989\times 10^{30}}{\left(2.998\times 10^8\right)^3}\right){}^{4/3}\right\},}\\
{\pi \times 8657.49422==\frac{96\left(6.674\times 10^{-11}\right)^{5/3}\left(m_1\times 1.989\times 10^{30}\right)\left(m_2\times 1.989\times
10^{30}\right)(\pi \times 116.6440307)^{11/3}}{5\times \left(2.998\times 10^8\right)^5\left(m_1 \times 1.989\times 10^{30}+m_2\times 1.989\times
10^{30}\right){}^{1/3}}}\\
{\times \left\{1-\left(\frac{743}{336}+\frac{11\times m_1\times m_2}{4\left(m_1+m_2\right){}^2}\right)\left(\frac{6.674\times 10^{-11}\pi \times
116.6440307\left(m_1+m_2\right)\times 1.989\times 10^{30}}{\left(2.998\times 10^8\right)^3}\right){}^{2/3}\right.}\\
{+4\pi \times \frac{6.674\times 10^{-11}\pi \times 116.6440307\left(m_1+m_2\right)\times 1.989\times 10^{30}}{\left(2.998\times 10^8\right)^3}}\\
{\left.+\left(\frac{34103}{18144}+\frac{13661\times m_1\times m_2}{2016\left(m_1+m_2\right){}^2}+\frac{59\times \left(m_1\times m_2\right){}^2}{\left(m_1+m_2\right){}^4}\right)\left(\frac{6.674\times
10^{-11}\pi \times 116.6440307\left(m_1+m_2\right)\times 1.989\times 10^{30}}{\left(2.998\times 10^8\right)^3}\right){}^{4/3}\right\},}\\
{\pi \times 418.0919129==\frac{96\left(6.674\times 10^{-11}\right)^{5/3}\left(m_1\times 1.989\times 10^{30}\right)\left(m_2\times 1.989\times
10^{30}\right)(\pi \times 42.73112738)^{11/3}}{5\times \left(2.998\times 10^8\right)^5\left(m_1 \times 1.989\times 10^{30}+m_2\times 1.989\times
10^{30}\right){}^{1/3}}}\\
{\times \left\{1-\left(\frac{743}{336}+\frac{11\times m_1\times m_2}{4\left(m_1+m_2\right){}^2}\right)\left(\frac{6.674\times 10^{-11}\pi \times
42.73112738\left(m_1+m_2\right)\times 1.989\times 10^{30}}{\left(2.998\times 10^8\right)^3}\right){}^{2/3}\right.}\\
{+4\pi \times \frac{6.674\times 10^{-11}\pi \times 42.73112738\left(m_1+m_2\right)\times 1.989\times 10^{30}}{\left(2.998\times 10^8\right)^3}}\\
{\left.+\left(\frac{34103}{18144}+\frac{13661\times m_1\times m_2}{2016\left(m_1+m_2\right){}^2}+\frac{59\times \left(m_1\times m_2\right){}^2}{\left(m_1+m_2\right){}^4}\right)\left(\frac{6.674\times
10^{-11}\pi \times 42.73112738\left(m_1+m_2\right)\times 1.989\times 10^{30}}{\left(2.998\times 10^8\right)^3}\right){}^{4/3}\right\},}\\
{\pi \times 764.1961764==\frac{96\left(6.674\times 10^{-11}\right)^{5/3}\left(m_1\times 1.989\times 10^{30}\right)\left(m_2\times 1.989\times
10^{30}\right)(\pi \times 53.04553744)^{11/3}}{5\times \left(2.998\times 10^8\right)^5\left(m_1 \times 1.989\times 10^{30}+m_2\times 1.989\times
10^{30}\right){}^{1/3}}}\\
{\times \left\{1-\left(\frac{743}{336}+\frac{11\times m_1\times m_2}{4\left(m_1+m_2\right){}^2}\right)\left(\frac{6.674\times 10^{-11}\pi \times
53.04553744\left(m_1+m_2\right)\times 1.989\times 10^{30}}{\left(2.998\times 10^8\right)^3}\right){}^{2/3}\right.}\\
{+4\pi \times \frac{6.674\times 10^{-11}\pi \times 53.04553744\left(m_1+m_2\right)\times 1.989\times 10^{30}}{\left(2.998\times 10^8\right)^3}}\\
{\left.+\left(\frac{34103}{18144}+\frac{13661\times m_1\times m_2}{2016\left(m_1+m_2\right){}^2}+\frac{59\times \left(m_1\times m_2\right){}^2}{\left(m_1+m_2\right){}^4}\right)\left(\frac{6.674\times
10^{-11}\pi \times 53.04553744\left(m_1+m_2\right)\times 1.989\times 10^{30}}{\left(2.998\times 10^8\right)^3}\right){}^{4/3}\right\},}\\
{\pi \times 1560.827218==\frac{96\left(6.674\times 10^{-11}\right)^{5/3}\left(m_1\times 1.989\times 10^{30}\right)\left(m_2\times 1.989\times
10^{30}\right)(\pi \times 68.32265222)^{11/3}}{5\times \left(2.998\times 10^8\right)^5\left(m_1 \times 1.989\times 10^{30}+m_2\times 1.989\times
10^{30}\right){}^{1/3}}}\\
{\times \left\{1-\left(\frac{743}{336}+\frac{11\times m_1\times m_2}{4\left(m_1+m_2\right){}^2}\right)\left(\frac{6.674\times 10^{-11}\pi \times
68.32265222\left(m_1+m_2\right)\times 1.989\times 10^{30}}{\left(2.998\times 10^8\right)^3}\right){}^{2/3}\right.}\\
{+4\pi \times \frac{6.674\times 10^{-11}\pi \times 68.32265222\left(m_1+m_2\right)\times 1.989\times 10^{30}}{\left(2.998\times 10^8\right)^3}}\\
{\left.+\left(\frac{34103}{18144}+\frac{13661\times m_1\times m_2}{2016\left(m_1+m_2\right){}^2}+\frac{59\times \left(m_1\times m_2\right){}^2}{\left(m_1+m_2\right){}^4}\right)\left(\frac{6.674\times
10^{-11}\pi \times 68.32265222\left(m_1+m_2\right)\times 1.989\times 10^{30}}{\left(2.998\times 10^8\right)^3}\right){}^{4/3}\right\},}\\
{\pi \times 3755.061951==\frac{96\left(6.674\times 10^{-11}\right)^{5/3}\left(m_1\times 1.989\times 10^{30}\right)\left(m_2\times 1.989\times
10^{30}\right)(\pi \times 92.72359945)^{11/3}}{5\times \left(2.998\times 10^8\right)^5\left(m_1 \times 1.989\times 10^{30}+m_2\times 1.989\times
10^{30}\right){}^{1/3}}}\\
{\times \left\{1-\left(\frac{743}{336}+\frac{11\times m_1\times m_2}{4\left(m_1+m_2\right){}^2}\right)\left(\frac{6.674\times 10^{-11}\pi \times
92.72359945\left(m_1+m_2\right)\times 1.989\times 10^{30}}{\left(2.998\times 10^8\right)^3}\right){}^{2/3}\right.}\\
{+4\pi \times \frac{6.674\times 10^{-11}\pi \times 92.72359945\left(m_1+m_2\right)\times 1.989\times 10^{30}}{\left(2.998\times 10^8\right)^3}}\\
{\left.+\left(\frac{34103}{18144}+\frac{13661\times m_1\times m_2}{2016\left(m_1+m_2\right){}^2}+\frac{59\times \left(m_1\times m_2\right){}^2}{\left(m_1+m_2\right){}^4}\right)\left(\frac{6.674\times
10^{-11}\pi \times 92.72359945\left(m_1+m_2\right)\times 1.989\times 10^{30}}{\left(2.998\times 10^8\right)^3}\right){}^{4/3}\right\},}\\
{\pi \times 11831.23042==\frac{96\left(6.674\times 10^{-11}\right)^{5/3}\left(m_1\times 1.989\times 10^{30}\right)\left(m_2\times 1.989\times
10^{30}\right)(\pi \times 136.3582345)^{11/3}}{5\times \left(2.998\times 10^8\right)^5\left(m_1 \times 1.989\times 10^{30}+m_2\times 1.989\times
10^{30}\right){}^{1/3}}}\\
{\times \left\{1-\left(\frac{743}{336}+\frac{11\times m_1\times m_2}{4\left(m_1+m_2\right){}^2}\right)\left(\frac{6.674\times 10^{-11}\pi \times
136.3582345\left(m_1+m_2\right)\times 1.989\times 10^{30}}{\left(2.998\times 10^8\right)^3}\right){}^{2/3}\right.}\\
{+4\pi \times \frac{6.674\times 10^{-11}\pi \times 136.3582345\left(m_1+m_2\right)\times 1.989\times 10^{30}}{\left(2.998\times 10^8\right)^3}}\\
{\left.\left.+\left(\frac{34103}{18144}+\frac{13661\times m_1\times m_2}{2016\left(m_1+m_2\right){}^2}+\frac{59\times \left(m_1\times m_2\right){}^2}{\left(m_1+m_2\right){}^4}\right)\left(\frac{6.674\times
10^{-11}\pi \times 136.3582345\left(m_1+m_2\right)\times 1.989\times 10^{30}}{\left(2.998\times 10^8\right)^3}\right){}^{4/3}\right\}\right\},}\\
{\left\{m_1,6,130\right\},\left\{m_2,6,130\right\},\text{GridLinesStyle}\to \text{Directive}[\text{Dashed}],\text{Axes}\to \text{False},}\\
{\text{GridLinesStyle}\to \text{Directive}[\text{Dashed}], \text{PlotLegends}\to \text{Placed}[\{\text{{``}Peak(45.376Hz,559.200Hz/s){''}},}\\
{\text{{``}Peak(58.445Hz,1142.134Hz/s){''}},\text{{``}Peak(79.318Hz,2747.764Hz/s){''}}, \text{{``}Peak(116.644Hz,8657.494Hz/s){''}},}\\
{\text{{``}Trough(42.731Hz,418.092Hz/s){''}},\text{{``}Trough(53.046Hz,764.196Hz/s){''}},\text{$\texttt{"}$Trough(68.323Hz,1560.827}\text{Hz}^2\text{)$\texttt{"}$},}\\
{\left.\left. \text{$\texttt{"}$Trough(92.724Hz,755.062}\text{Hz}^2\text{)$\texttt{"}$},\text{$\texttt{"}$Trough(136.358Hz,11831.2}\text{Hz}^2\text{)$\texttt{"}$}\right\},\{0.76,0.60\}\right],}\\
{\text{Ticks}\to \{\text{Range}[-10,10,2]\},\text{ContourStyle}\to \{\{\text{Hue}[0.1],\text{AbsoluteThickness}[3]\},}\\
{\{\text{Hue}[0.2],\text{AbsoluteThickness}[3]\},\{\text{Hue}[0.3],\text{AbsoluteThickness}[3]\},}\\
{\{\text{Hue}[0.4],\text{AbsoluteThickness}[3]\},\{\text{Hue}[0.5],\text{AbsoluteThickness}[3]\}, }\\
{\{\text{Hue}[0.60],\text{AbsoluteThickness}[3]\},\{\text{Hue}[0.7],\text{AbsoluteThickness}[3]\},}\\
{\{\text{Hue}[0.8],\text{AbsoluteThickness}[3]\},\{\text{Hue}[0.9],\text{AbsoluteThickness}[3]\}\},}\\
{\left.\text{FrameLabel}\to \left\{\texttt{"}m_1 \;({m_\odot})\texttt{"},\texttt{"}m_2 \;({m_\odot})\texttt{"}\right\},\text{AspectRatio}\to 0.86\right]}\)
\end{appendix}


\begin{thebibliography}{C}
\bibitem{Einstein:1915} Einstein, A. Die Feldgleichungun der Gravitation. \textit{Sitzungsber. K. Preuss. Akad. Wiss.}, 844-847 (1915).
\bibitem{Einstein:1916} Einstein, A. The foundation of the general theory of relativity. \textit{Annalen Phys.} \textbf{14}, 769-822 (1916).
\bibitem{Einstein:1917} Einstein, A. N\"{a}herungsweise Integration der Feldgleichungen der Gravitation. \textit{Sitzungsber. K. Preuss. Akad. Wiss.}, 688-696 (1916).
\bibitem{Schwarzschild:1916} Schwarzschild, K. \"{U}ber das Gravitationsfeld eines Massenpunktes nach der Einsteinschen Theorie. \textit{Sitzungsber. K. Preuss. Akad. Wiss.}, 189-196 (1916).
\bibitem{Schwarzschild:1917} Schwarzschild, K. \"{U}ber das Gravitationsfeld eines Massenpunktes nach der Einsteinschen Theorie. \textit{Sitzungsber. K. Preuss. Akad. Wiss.}, 424-434 (1916).
\bibitem{Einstein:1918} Einstein, A. \"{U}ber Gravitationswellen. \textit{Sitzungsber. K. Preuss. Akad. Wiss.}, 154-167 (1918).
\bibitem{Kerr:1963} Kerr, R. P. Gravitational field of a spinning mass as an example of algebraically special metrics. \textit{Physical review letters} \textbf{11}, 237-238 (1963).
\bibitem{Weber:1961} Weber, J. {\&} Mcvittie, G. C. \textit{General Relativity and Gravitational Waves}. (Interscience Publishers, Inc, 1961).
\bibitem{Abramovici:1992} Abramovici, A.\textit{ et al.} LIGO: The laser interferometer gravitational-wave observatory. \textit{Science} \textbf{256}, 325-333 (1992).
\bibitem{Harry:2010} Harry, G. M. {\&} Collaboration, L. S. Advanced LIGO: the next generation of gravitational wave detectors. \textit{Classical and Quantum Gravity} \textbf{27}, 084006 (2010).
\bibitem{Baker:2001} Baker, J., Br\"{u}gmann, B., Campanelli, M., Lousto, C. O. {\&} Takahashi, R. Plunge waveforms from inspiralling binary black holes. \textit{Physical review letters} \textbf{87}, 121103 (2001).
\bibitem{Damour:2008} Damour, T., Nagar, A., Hannam, M., Husa, S. {\&} Br\"{u}gmann, B. Accurate effective-one-body waveforms of inspiralling and coalescing black-hole binaries. \textit{Physical Review D} \textbf{78}, 044039 (2008).
\bibitem{Hulse:1974} Hulse, R. {\&} Taylor, J. A high-sensitivity pulsar survey. \textit{The Astrophysical Journal} \textbf{191}, L59 (1974).
\bibitem{Blanchet:1995} Blanchet, L., Damour, T., Iyer, B. R., Will, C. M. {\&} Wiseman, A. G. Gravitational-radiation damping of compact binary systems to second post-Newtonian order. \textit{Physical Review Letters} \textbf{74}, 3515 (1995).
\bibitem{Abbott:2016} Abbott, B. P.\textit{ et al.} Observation of gravitational waves from a binary black hole merger. \textit{Physical review letters} \textbf{116}, 061102 (2016).
\bibitem{Abbott:2017} Abbott, B.\textit{ et al.} Localization and broadband follow-up of the gravitational-wave transient GW150914. \textit{The Astrophysical journal letters} \textbf{826}, L13 (2016).
\bibitem{Abbott:2018} Abbott, B.\textit{ et al.} GW151226: Observation of gravitational waves from a 22-solar-mass binary black hole coalescence. \textit{Physical Review Letters} \textbf{116}, 241103 (2016).
\bibitem{Scientific:2017} Scientific, L.\textit{ et al.} GW170104: observation of a 50-solar-mass binary black hole coalescence at redshift 0.2. \textit{Physical Review Letters} \textbf{118}, 221101 (2017).
\bibitem{Abbott:2019} Abbott, B. P.\textit{ et al.} GW170814: a three-detector observation of gravitational waves from a binary black hole coalescence. \textit{Physical review letters} \textbf{119}, 141101 (2017).
\bibitem{Abbott:2020} Abbott, B. P.\textit{ et al.} GW170817: observation of gravitational waves from a binary neutron star inspiral. \textit{Physical Review Letters} \textbf{119}, 161101 (2017).
\bibitem{https:1} https://losc.ligo.org/events/GW150914/. LIGO Gravitational Wave Strain Data.
\bibitem{Sahoo:1998} Sahoo, P. {\&} Riedel, T. \textit{Mean value theorems and functional equations}. (World Scientific, 1998).
\bibitem{Hawking:1972} Hawking, S. W. Black holes in general relativity. \textit{Communications in Mathematical Physics} \textbf{25}, 152-166 (1972).

\end{thebibliography}
\end{document}